\documentclass[useAMS,usenatbib]{mn2e}

\usepackage{amsmath}
\usepackage{amsfonts}
\usepackage{times}
\usepackage{amssymb}
\usepackage{makeidx}
\usepackage{amssymb}
\usepackage{graphicx}
\usepackage{url}

\title [The mid-IR environments of MMB  Masers]{The mid-infrared environments of  6.7 GHz Methanol Masers from the Methanol Multi-Beam Survey}

\author[M Gallaway et al.]
{M. Gallaway,$^1$\thanks{Email:M.Gallaway@herts.ac.uk}
 M.~A.~Thompson,$^1$ P.~W.~Lucas,$^1$ G.~A.~Fuller,$^2$ J.~L.~Caswell,$^3$
 \newauthor 
J.~A.~Green,$^{2,3}$
M.~A.~Voronkov,$^{3,4}$
S.~L.~Breen,$^{3}$ 
L.~Quinn,$^{2}$
S.~P.~Ellingsen,$^5$
\newauthor
A.~Avison,$^2$
D.~Ward-Thompson,$^6$
J.~Cox$^6$
\\
$^1$ Centre for Astrophysics Research, University of Hertfordshire, College Lane, Hatfield, Hertfordshire, AL10 9AB, UK;\\
$^2$ Jodrell Bank Centre for Astrophysics, Alan Turing Building, School of Physics and Astronomy, University of Manchester, Manchester, M13
9PL, UK;\\
$^3$ CSIRO Astronomy \& Space Science, Australia Telescope National Facility, PO Box 76, Epping, NSW, Australia, 2121;\\
$^4$ Astro Space Centre, Profsouznaya st. 84/32, 117997 Moscow, Russia;\\
$^5$ School of Mathematics and Physics, University of Tasmania, Private Bag 37, Hobart, Tasmania 7001, Australia;\\
$^6$ Department of Physics and Astronomy, Cardiff University, 5 The Parade, Cardiff, CF24 3YB, UK;\\
}
\begin{document}

\date{}

\pagerange{\pageref{firstpage}--\pageref{lastpage}} \pubyear{2011}

\maketitle

\label{firstpage}

\begin{abstract}
We present a study of the mid-infrared environments and association with star formation tracers of 6.7 GHz methanol masers taken from the Methanol Multi-Beam (MMB) Survey. Our ultimate goal is to establish the mass of the host star and its evolutionary stage for each maser site. As a first step, the GLIMPSE survey of the Galactic Plane is utilised to investigate the environment of 776 methanol masers and we find that while the majority of the masers are associated with mid-infrared counterparts, a significant fraction (17\%) are not associated with any detectable mid-infrared emission. A number of the maser counterparts are clearly extended with respect to the GLIMPSE point spread function and we implement an adaptive non-circular aperture photometry (ANCAP) technique to determine the fluxes of the maser counterparts. The ANCAP technique doubles the number of masers with flux information at all four wavelengths compared to the number of the corresponding counterparts obtained from the GLIMPSE Point Source Catalogue. The colours of the maser counterparts are found to be very similar to the smaller study carried out by \citet{Ell06}. The MMB masers are weakly associated with Extended Green Objects (EGOs) and Red MSX Survey (RMS) embedded sources (YSO and HII classifications) with 18\% and 12\% of masers associated with these objects respectively. 
The majority of MMB masers (60\%) have detectable GLIMPSE infrared counterparts  but have \emph{not} been identified with previously recognised tracers of  massive star formation;  this confirms  that  the MMB survey has the potential to identify massive star forming regions independent of infrared selection.
\end{abstract}


\section{Introduction}
\label{sect:intro}

Massive stars (M$>8M_\odot$) are a key component in the evolution of
the Interstellar Medium (ISM). They provide the bulk of the radiation that ionizes the ISM
 and transfer kinetic energy into the ISM via stellar winds, outflows and
supernovae. This injection of energy may trigger new generations of star
formation \citep[e.g.][]{elmegreen1992}. Yet our understanding of the formation of  massive stars is
very limited in contrast to low mass star formation. This is due in
part to the rarity of high mass stars but also to their large distances from the Sun (typically $>$ 1 kpc) and their very brief
pre-main sequence lives. Massive stars enter the main sequence before accretion has finished and therefore whilst still deeply
embedded in an obscuring dusty nebula \citep{garay1999,Zin2007}. Due to the observational difficulty in identifying the brief and rare early stages of massive star formation, it has been only within the last decade that massive young stellar objects (MYSOs) and high mass ``protostellar objects'' (HMPOs\footnote{as these objects are likely hydrogen-burning they should strictly not be referred to as ``protostellar'' but we reproduce here the original acronym of \citet{beuther2002} to avoid confusion.}) have been identified in large numbers
\citep[e.g.][]{Lum02, beuther2002,urquhart2011}.

Despite these limitations a tentative evolutionary sequence for  massive star formation 
has emerged. The sequence commences with a sub-millimetre bright cold core with a typical diameter of $<0.5$ pc, a mass
of $10^2 - 10^3 M _\odot$ and a typical temperature of 10-20\;K,
embedded within a much larger molecular clump $\sim$1000 M$_{\odot}$ which in turn is embedded within a giant
molecular cloud. The presence of an Infrared Dark Cloud (IRDC)  is often observed at this point
\citep{Simon06a,Peretto09,kauffmann2010,peretto2010a,peretto2010b}, but the detection of an IRDC is dependent upon observing the cloud against a diffuse mid-infrared background.  The core collapses into a hydrostatically supported optically thick protostellar embryo which then accretes material from the core, most likely via a circumstellar disk. The onset of the main sequence does not halt the accretion, which continues until the young massive star is hot enough to produce ionizing UV
photons. These photons ionize the surrounding material, creating an HII
region. Initially the HII region is gravitationally bound
\citep{Keto03} but as the ionizing flux rises the surrounding hot core is unable
to contain it gravitationally.

At this point the inflow of material
becomes important in quenching the expansion of the HII region. The
ionized gas is contained as a $\leq$ 10000 AU diameter bubble
\citep{kurtz2005} and is observed in the radio waveband as a Hyper
Compact HII (HCHII) region. During this phase the material trapped
within the HCHII region continues to accrete. Eventually a combination
of reduced inflow and increased UV flux results in the HCHII
expanding to form an Ultra Compact HII (UC HII)
region \citep{Hoare07} and as a consequence any further accretion may be
terminated. The UCHII region continues to expand, with the rate of
expansion and final size and morphology being determined by the mass
of the driving core and the density of the surrounding ambient
material \citep{ellingsen2005}. Finally, the ionizing source becomes visible as a main
sequence massive star \citep[][and references therein]{Hoare07, Purcell, Keto03,Motte08,Wood89}.
\paragraph*{}

One of the principal tracers of massive star formation is the
6.7\;GHz Class II methanol maser \citep{menten1991}, thought to be exclusively associated with massive star formation 
\citep{Min03,Cragg05,Ell06,Walsh99,green2012} and to primarily trace the stage between the infrared-dark phase and the UC HII region \citep{Walt05,Ell06}. Whilst much work has been recently carried out on the evolutionary period of star formation traced by the Class II methanol masers \citep[e.g.][]{Breen2011,breen2011b,ellingsen2011, breen2010b,breen2010a} a number of details remain unclear, particularly the minimum bolometric luminosity traced by the masers \citep{Min03,xu2008} and the relationship between the masers and their wider infrared environment. In part this is due to the historical lack of an unbiased and  large sample of masers with positions sufficiently precise to allow reliable identification with infrared and sub-mm counterparts. Many of the presently known methanol masers were discovered via
large-beam  single dish observations of IRAS point sources (see \citealt{Pesta07} for a compilation of detected masers), hence  are both biased towards bright far-infrared  sources and do not have positions measured to an accuracy of better
than an arcminute or so.  Most of the infrared/maser comparison studies that have been carried out so far are limited to small samples of masers whose positions have been measured interferometrically or inferred by other means \citep[e.g.][]{Ell06,pandian2010}.

These constraints have now been 
 addressed by the Methanol Maser Multi-Beam (MMB)
Survey: a deep, high resolution, untargeted survey of the Galactic
Plane and Magellanic Clouds with the goal of providing the first
comprehensive catalogue of Class II 6.7\;GHz methanol masers
\citep{Green09,Caswell2010,green2010,caswell2011,green2012}.  The 6.7 GHz methanol masers were first found using a multi-beam receiver on Parkes and then individually re-observed with the ATCA and Merlin interferometers to obtain sub-arcsecond positions \citep[see][for further details of the survey strategy]{Green09}. The Parkes MMB survey region encompasses $-174 < l < 60$ and $|b|<2$ \citep{Green09}, with  plans to later extend the survey  to the  Northern Galactic Plane.  With this two step approach the MMB Survey will provide an essentially \emph{complete} survey of 6.7 GHz methanol masers in the Galaxy, each with positional accuracy suitable for follow-up comparison studies in the infrared and sub-mm. 

In this paper we describe a follow-up study which has been carried out with an interim  MMB catalogue of 824 6.7 GHz masers  and the GLIMPSE (Galactic Legacy Infrared Mid-Plane Survey Extraordinaire) infrared survey of the Galactic Plane \citep{Ben03}.  The GLIMPSE survey (including the individual I, II and 3D surveys) covers $-65\le l \le65$ and approximately $|b|\le 1$ in 4 infrared bands  (3.6, 4.5, 5.8 and 8.0\;$\mu$m respectively and usually referred to as Bands 1 to 4 in that order) with the IRAC camera on \emph{Spitzer}.

Our interim catalogue is derived from those masers that had been interferometrically positioned at the time this work was carried out. The interim catalogue includes 684 masers in the longitude range $186\degr \le l \le 20\degr$  that have already been published in \cite{Caswell2009}, \cite{Caswell2010}, \cite{green2010}, \cite{caswell2011} and \cite{green2012}\footnote{The MMB catalogue is available at \texttt{http://astromasers.org}}.  The remaining 140 masers lie in the longitude range $20\degr \le l \le 186\degr$ and will be reported in a further publication (Fuller et al., 2012, in prep). Note that our interim catalogue does not include 23 masers from the published catalogue whose positions had not yet been interferometrically determined at the time this work was carried out.

Using GLIMPSE we seek to identify the infrared counterparts of the masers so that we may investigate both their environment (i.e.~are the masers preferentially found towards infrared dark regions such as IRDCs?), the colours of the maser counterparts and the likely nature of their exciting sources. This work builds upon that of \cite{Ell06} but with a much larger sample of masers and an aperture photometry approach that is not biased to the limitations of the GLIMPSE point source catalogue in regions of crowding or high background. Future publications will examine the spectral energy distributions and infrared luminosities of the masers (using MIPSGAL, Hi-GAL and ATLASGAL data; \citealt{carey2009,molinari2010a,schuller2009}), and their association with other tracers of massive star formation. Preliminary work in characterising the nature of the MMB masers is included in the present paper from comparisons with the Red MSX Source (RMS) survey \citep{urquhart2011,urquhart2008,Lum02}.

\section{Method}
\label{sect:method}

The process of comparison with the GLIMPSE survey was performed in two ways: a detailed inspection of GLIMPSE cut-out images centred on each maser and a positional cross-match of the maser coordinates with the GLIMPSE point source catalogues, similar to that performed by \citet{Ell06}. The first method allows a detailed investigation of the appearance of the maser counterparts  and the wider environments in which the masers are found, whereas the second method permits the study of the infrared colours of a large sample of masers. We use the GLIMPSE I, II \& 3D survey data and images for maximum areal coverage.

The GLIMPSE Point Source Catalogue contains 69.7 million sources extracted from the GLIMPSE I, II \& 3D images which are considered to be high reliability ($\ge$99.5\%) point sources \citep{churchwell2009}. A less reliable but more complete GLIMPSE Point Source Archive (GPSA) contains some 104 million point sources. In order to be included in the GPSC each source must be a point source that is detected at least twice in one wavelength band and once in an adjacent wavelength band (referred to as the ``2+1'' criteria) with a signal-to-noise of greater than 5 for both detections. The criteria for the remaining bands are relaxed to be either $\ge 3\sigma$ or upper limits. For the GPSA the 2+1 criteria are relaxed to a detection $\ge 5\sigma$ twice in any one wavelength band or once in two adjacent wavelength bands, resulting in a more complete but less reliable catalogue of point sources.

We cross-matched the MMB catalogue to the GPSC and GPSA using a 2\arcsec\ matching radius (for discussion of this particular selection see Sect.~\ref{sect:ancap_method}). Within the GLIMPSE I, II \& 3D survey area  the MMB catalogue contains 776 6.7 GHz Class II methanol masers. When cross-matching against the GLIMPSE catalogues we find 430 of these masers with counterparts in the GPSC and 519 with counterparts in the GPSA (note that the GPSA contains all sources found in the GPSC cross-match, i.e. the GPSC is a more reliable subset of the GPSA). 

However, less than half of these sources have detections in all 4 wavelength bands (219 sources from the GPSC and 253 from the GPSA) and only $\sim$75\% have detections in any 3 GLIMPSE bands. This is an inevitable outcome of the GLIMPSE point source detection algorithm which, although highly successful in the detection of stellar point sources, may not be optimised for the crowded and complex environments of massive star fomation \citep[see Sect.~2.2 of][and references therein for further discussion]{robitaille2008}. Visual inspection of the GLIMPSE cut-out images confirmed that this is indeed the case; the majority of MMB masers were associated with emission in all 4 bands, in many cases with nebulous emission or emission extended beyond the \emph{Spitzer} PSF, that is not catalogued within the GPSC or GPSA.

As the primary aim of our work is to obtain accurate mid-infrared colours of the MMB masers we have implemented an adaptive non-circular aperture
photometry technique (referred to as ANCAP)  to measure the fluxes of masers in all four IRAC bands. In Sect.~\ref{sect:visual} we describe our visual inspection and classification methods, followed by our aperture photometry technique in Sect.~\ref{sect:ancap_method}. Finally we inspect the relationship between MMB masers and a number of other published catalogues of star formation tracers (e.g.~Red MSX Survey objects, \cite{urquhart2011}, and Extended Green Objects, \cite{Cyg08}) in Sect.~\ref{sect:cats}.

\subsection{Visual inspection of GLIMPSE images}
\label{sect:visual}

 GLIMPSE 12 arcminute FITS cut-out images were obtained for all four IRAC bands
 via the NASA/IPAC Gator
 database query tool, with each
 image centred on an MMB maser. We select images of this size as it is the typical angular size of many IRDC complexes \citep[e.g.][]{Peretto09,Simon06a} and star forming complexes/Giant Molecular clouds \citep[e.g.][]{solomon1987}. RGB images were generated by a script with
 IRAC bands 1, 2 and 4 (i.e.~3.6, 4.5 and 8 $\mu$m) represented by blue, green and red, respectively and
 overlaid with the MMB catalogue. The RGB images were used to classify
 the maser counterparts by environment and extension, in addition to the identification of masers associated with Extended Green Objects (EGOs), Infrared Dark Clouds (IRDCs) and infrared clusters. 
 
 IRDCs have high levels of extinction, often $A_{v}\geq100$, and appear as dark ``holes''
within the diffuse 8.0\;$\mu$m emission associated with Polycyclic
Aromatic Hydrocarbons (PAHs) \citep{Parsons09,Simon06a,Peretto09,Egan98}. We identify IRDCs in the GLIMPSE images by the absence of background 8.0\;$\mu$m emission and a reduction in the number of background stars. Given the often irregular and filamentary appearance of IRDCs \citep{Simon06a, Peretto09}, visual associations of masers and IRDCs are usually more accurate than simple nearest-neighbour matching \citep[see e.g.~][]{Parsons09} and so we choose the visual approach here.

For EGOs and infrared clusters our visual search is complementary to positional cross matches against the \cite{Cyg08} EGO and  \citep{Mercer, Bica, Froebrich} cluster catalogues. By visually examining the GLIMPSE cutouts we can search for associated EGOs and/or clusters that may have been missed in the catalogues. EGOs are identified by the presence of bright green extended emission in the cutout images, using a similar process to that employed by \cite{Cyg08}. Infrared clusters and cluster candidates are identified by inspecting  the  GLIMPSE images and, where available, UK Infrared Deep Sky Survey (UKIDSS) Galactic Plane Survey images \citep{Lucas08}. We identify clusters via local overdensities of infrared stars, taking as a lower limit the presence of at least 10 stars as cluster members, and with the maser lying within the projected bounds of the cluster.

\subsection{Adaptive Non-circular Aperture Photometry of the maser counterparts}
\label{sect:ancap_method}

We have implemented  an adaptive non-circular aperture
photometry (ANCAP) technique to measure the fluxes of extended infrared counterparts to the masers and provide as large a sample of masers with fluxes in all 4 GLIMPSE bands as possible. Our technique is based on that used by \cite{Cyg08} to measure the
fluxes of EGOs within the GLIMPSE image archive. The ``region'' feature of
the SAO DS9 FITS viewer was used to draw non-circular apertures around
our sources. The counterparts are identified by the presence of a co-located peak
in all four IRAC bands within 2 arcseconds of the maser, a radius selected to be consistent with previous studies by \cite{Ell06} and  the typical resolution of IRAC wavelength band 4 (8 $\mu$m). If a clear peak cannot be identified in all four bands within 2\arcsec\ of the maser we do not undertake photometry for this maser, to avoid including local regions of nebulosity which may encompass large areas and multiple shorter wavelength sources. By utilising all four
bands to identify the counterpart  the rate of misidentification is reduced. In the case where  there is more than one possible
counterpart we also make no measurement, to avoid confusion. 

\begin{figure*}
\includegraphics[scale=0.4]{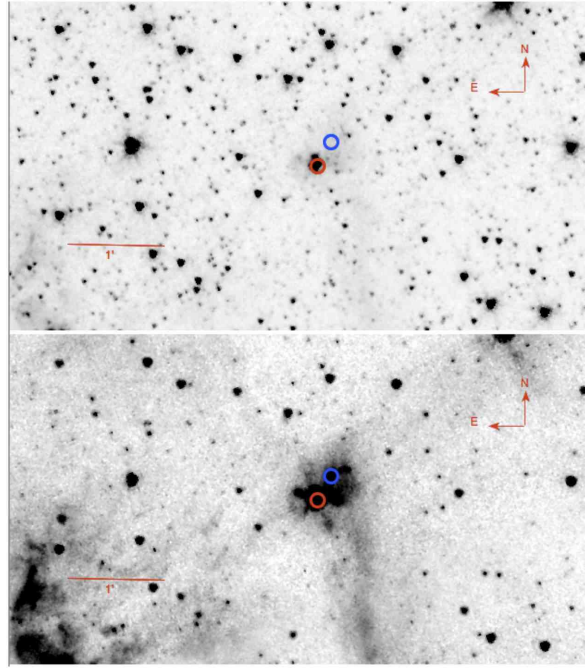} 
\caption{The above images show an illustration of the process used to select the infrared counterpart. The above images are Band 1 (3.6\;$\mu$m) top left, Band 2 (4.5\;$\mu$m) top right, Band 3 (5.8\;$\mu$m) bottom left and Band 4 (8.0\;$\mu$m) bottom right. The red and blue circles (dark grey and light grey in the print version), which are 2\arcsec\ in radius,  indicate maser positions.}
\label{Figure1}
\end{figure*}

Figure \ref{Figure1} illustrates the process used in identifying the infrared counterpart. In this
example there are two masers in the field, marked by the red and blue circles in each frame.  If we
used only the band 4 image to determine our association we would incorrectly associate both masers with the large ``double-lobed''
object. However by inspecting  the band 1 image we see that in fact only the maser marked with the red circle is associated with a source with flux in
all four bands. In our example, Figure \ref{Figure1} bands 3 and 4 would be rejected for aperture size and shape determination, due to the
likelihood of contamination. In this case band 2 would be used for this function.

\begin{figure*}
\includegraphics[scale=0.4]{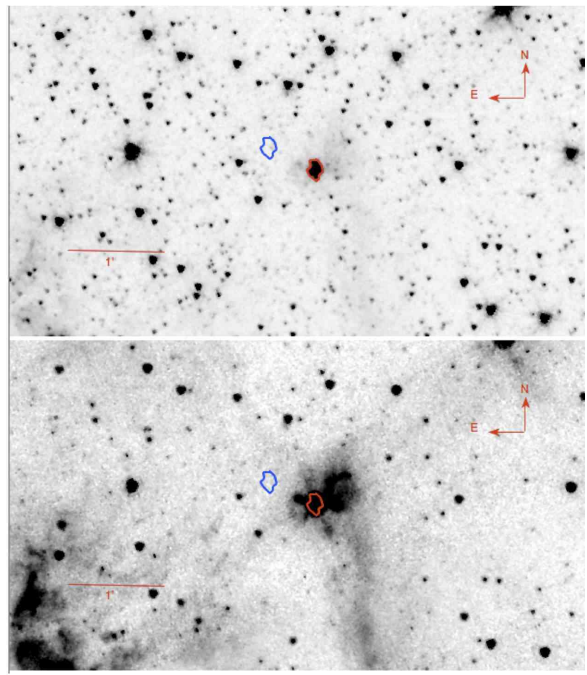} 
\caption{Example of the regions used in the non-circular aperture photometry. The above images are Band 1 (3.6\;$\mu$m) top left, Band 2 (4.5\;$\mu$m) top right, Band 3 (5.8\;$\mu$m) bottom left and Band 4 (8.0\;$\mu$m) bottom right. The vector in each image has a length of 1 arcminute. The  red and blue polygons (dark greay and light grey in the print version) are the apertures and background respectively.}
\label{Figure2}
\end{figure*}

The aperture shapes are selected to avoid contamination from field sources, whilst containing as much flux as possible. Preference is given to band 4 when choosing the aperture shape, although in cases where selecting the aperture in band 4 causes avoidable contamination band 3 is used instead.   The background is removed by subtracting the mean level contained in an identically sized aperture located in an area of the image representative of the background local to the maser source. Both aperture and background regions are constant in size,
shape and position across all four bands. The aperture sizes used are typically less than 20\arcsec\ in radius. The DS9 funtools plugin \url{(http://www.cfa.harvard.edu/~john/funtools/ds9.html)}  allows us
to integrate the flux in both regions and thereby obtain a raw flux to which we applied unit, scale and aperture corrections as outlined in
the IRAC cookbook \url{(http://ssc.spitzer.caltech.edu/postbcd/irac_reduction.html)}. Figure \ref{Figure2} shows a set of IRAC images overlaid with the region which make up the aperture and the background.

\begin{figure*}
\begin{center}
\includegraphics[scale=0.35]{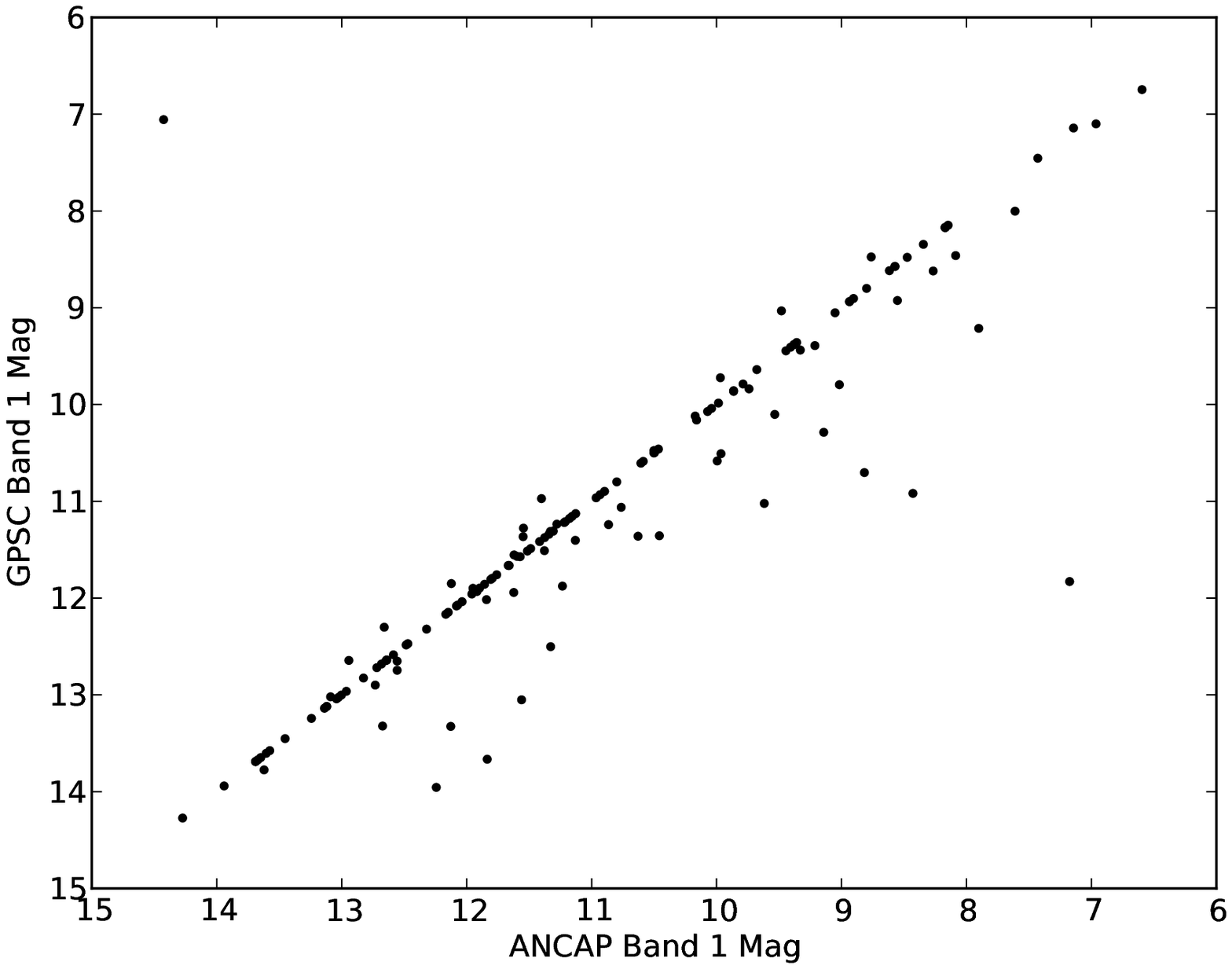} \includegraphics[scale=0.35]{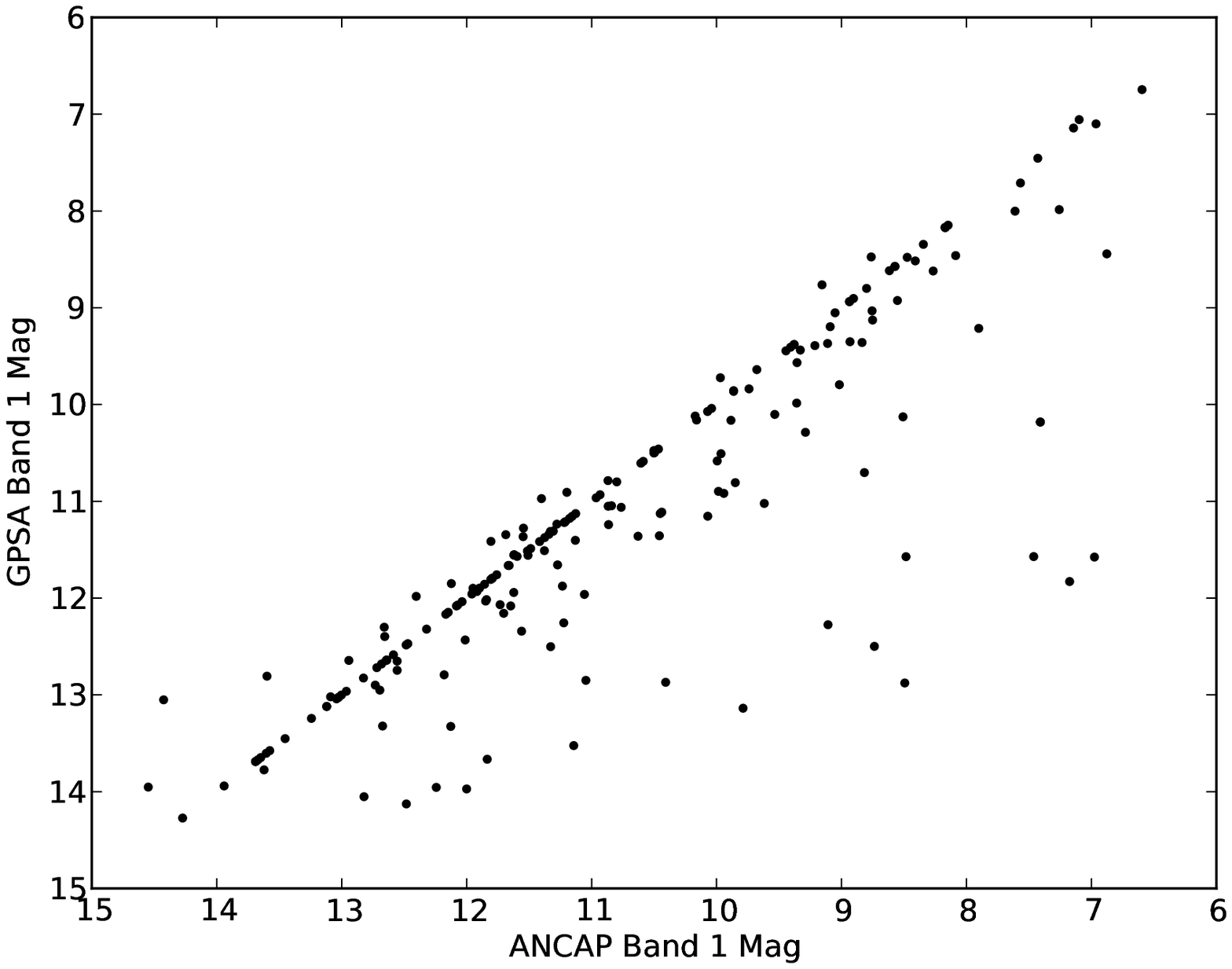} \\
\includegraphics[scale=0.35]{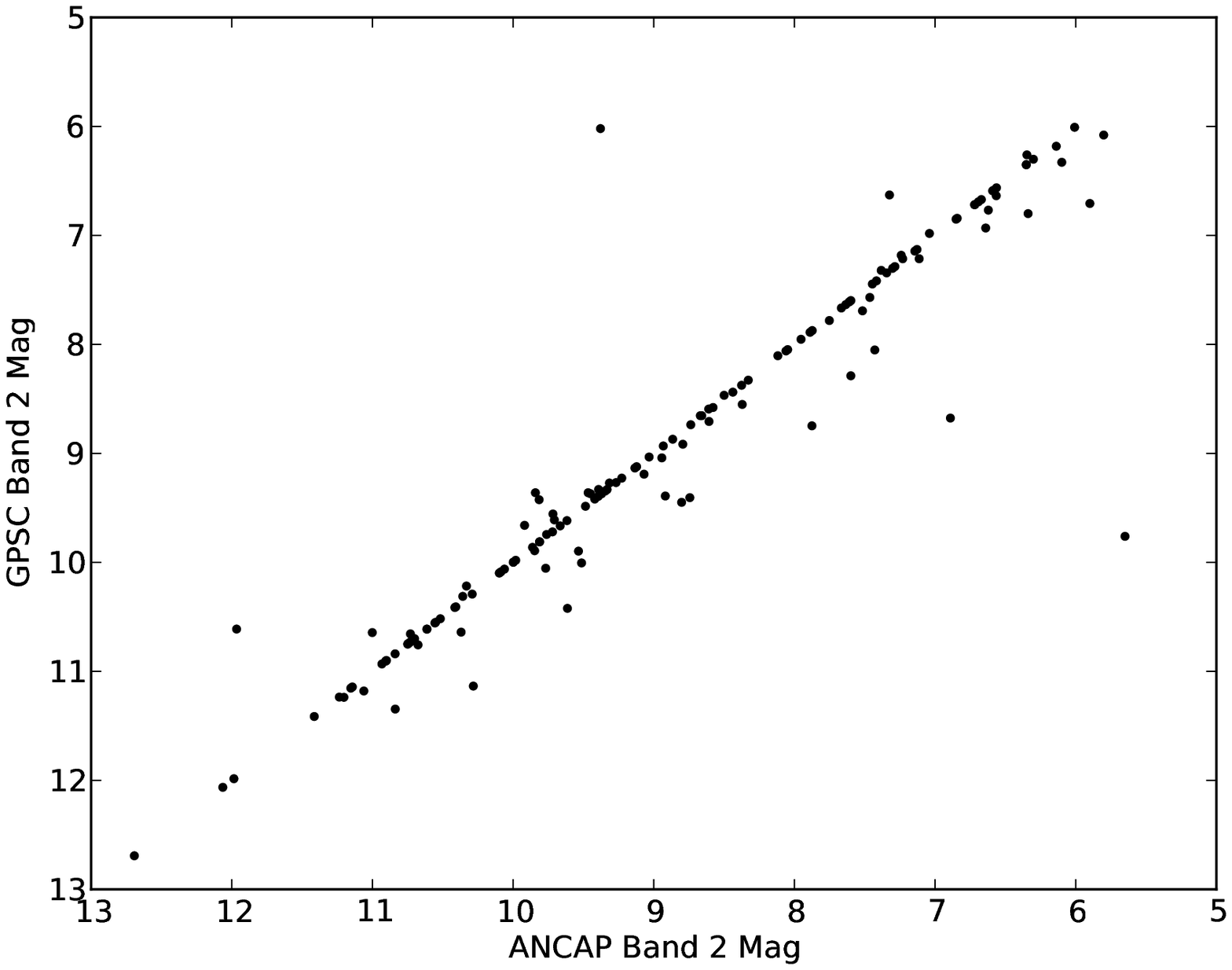} \includegraphics[scale=0.35]{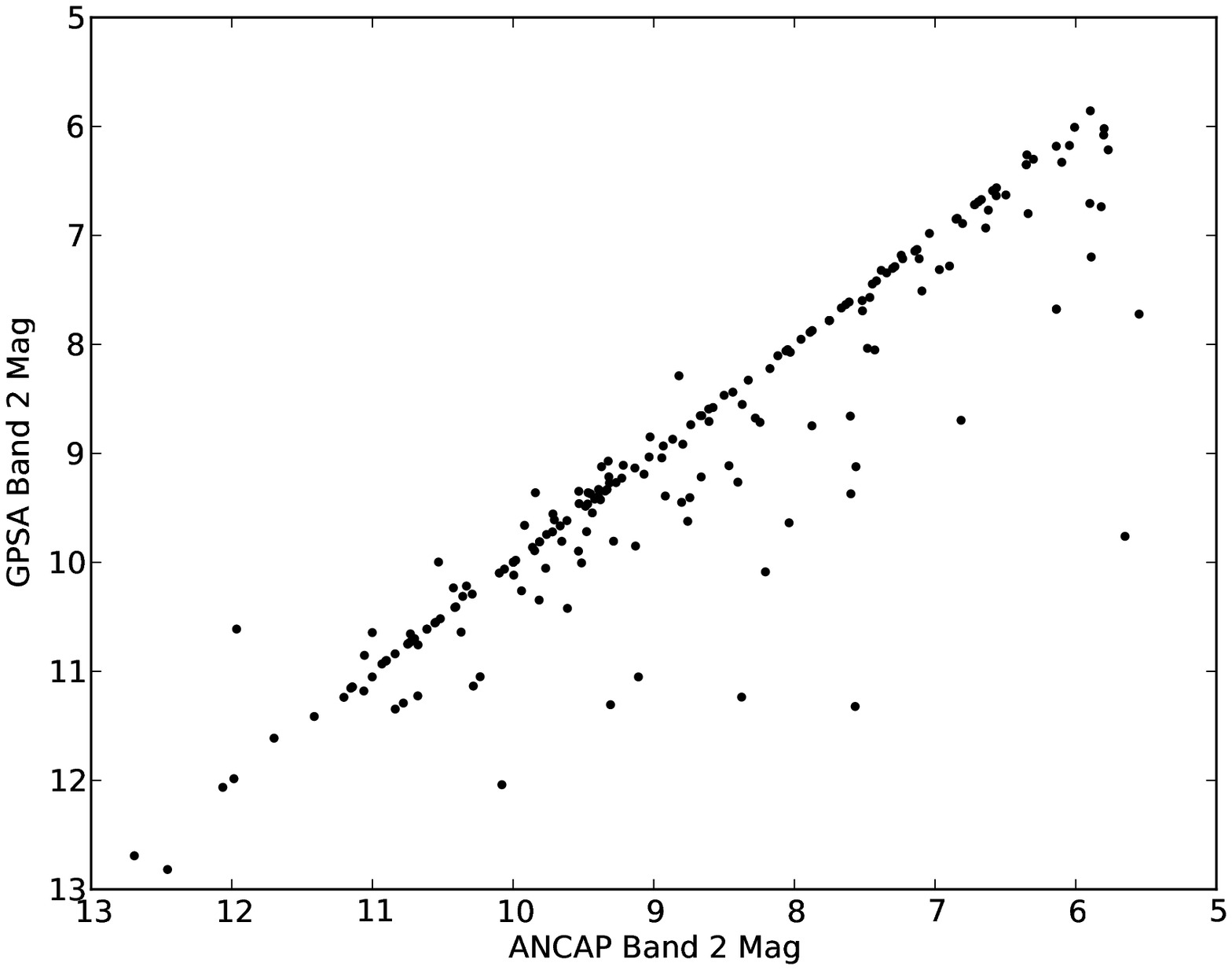} \\
\includegraphics[scale=0.35]{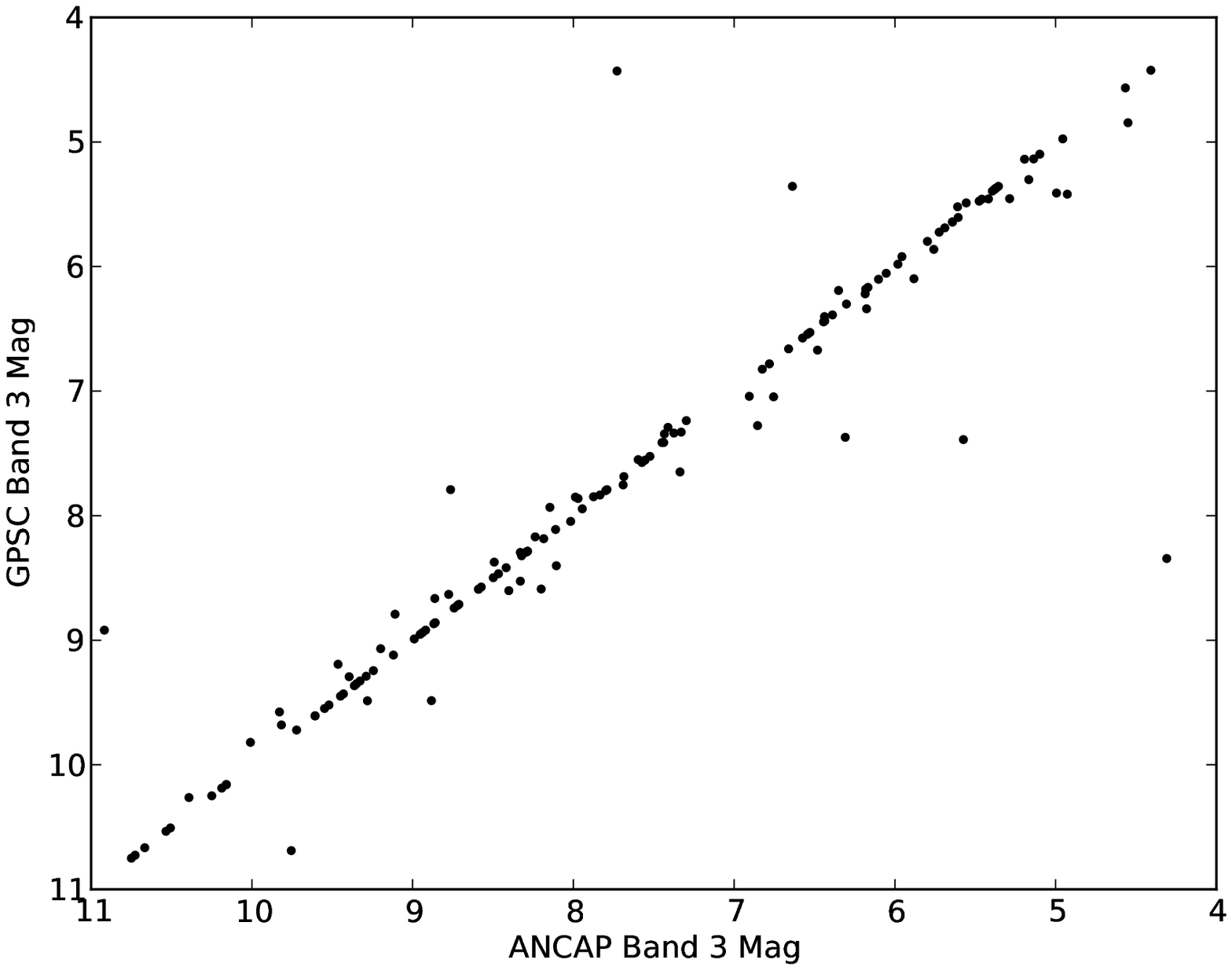} \includegraphics[scale=0.35]{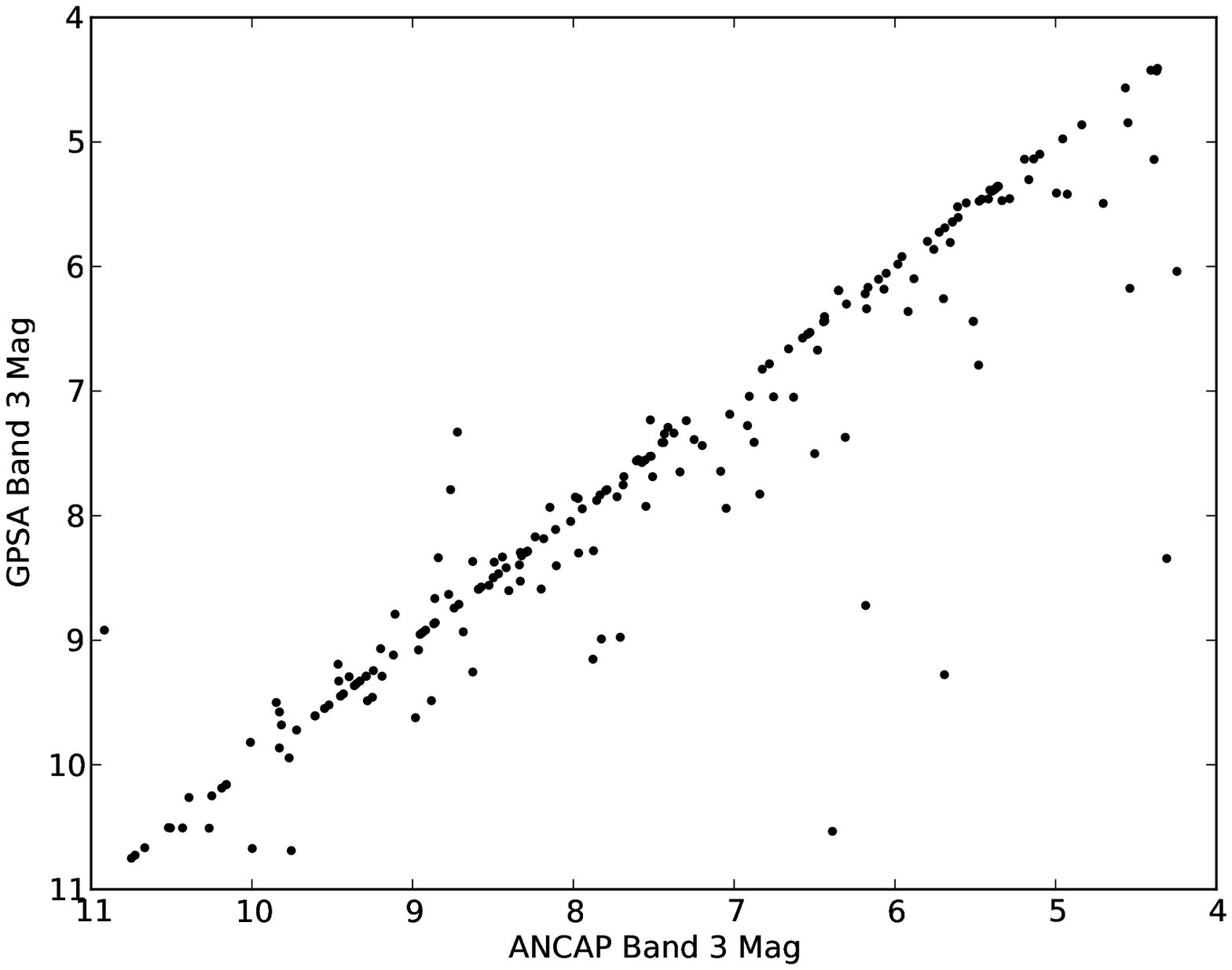} \\
\includegraphics[scale=0.35]{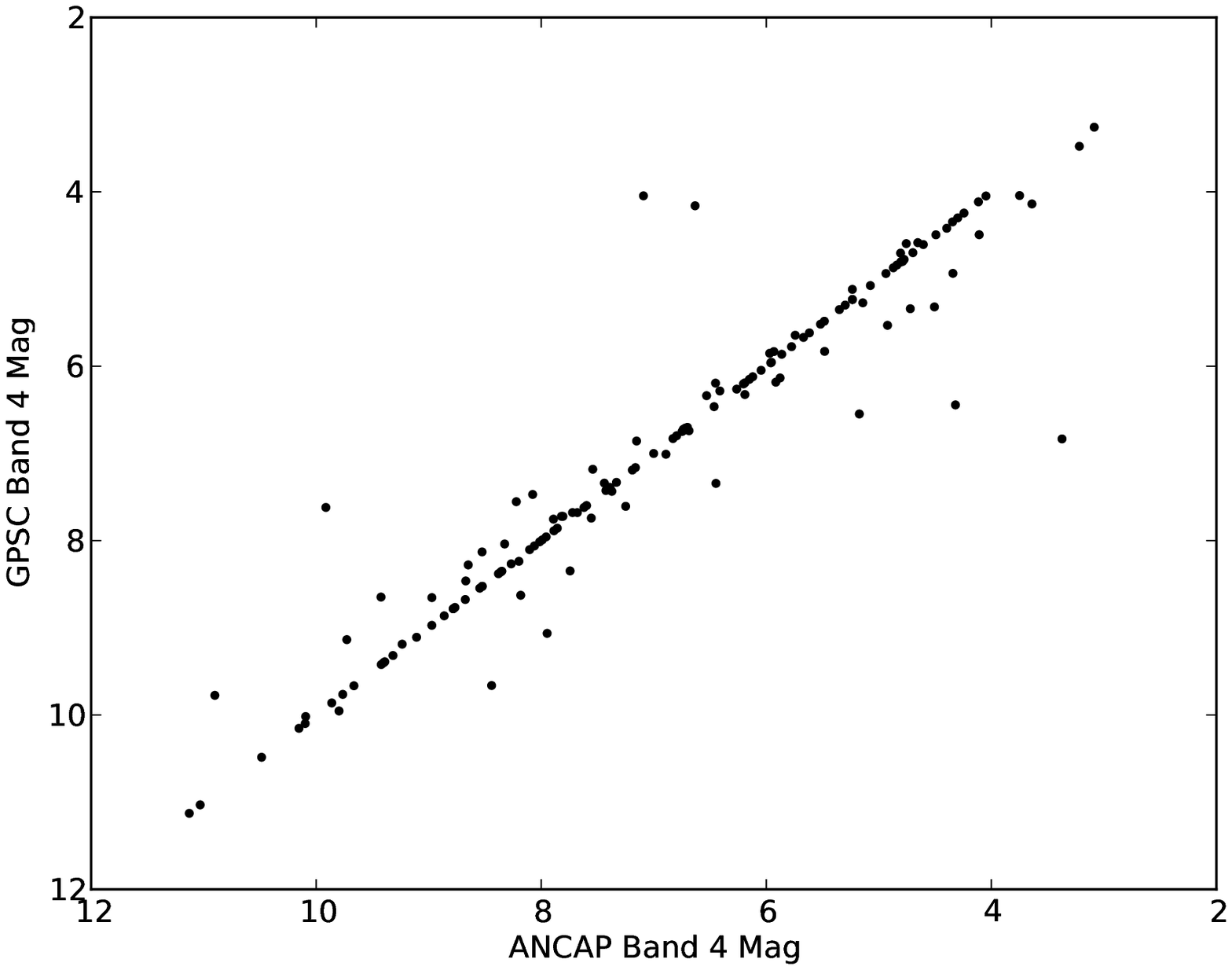} \includegraphics[scale=0.35]{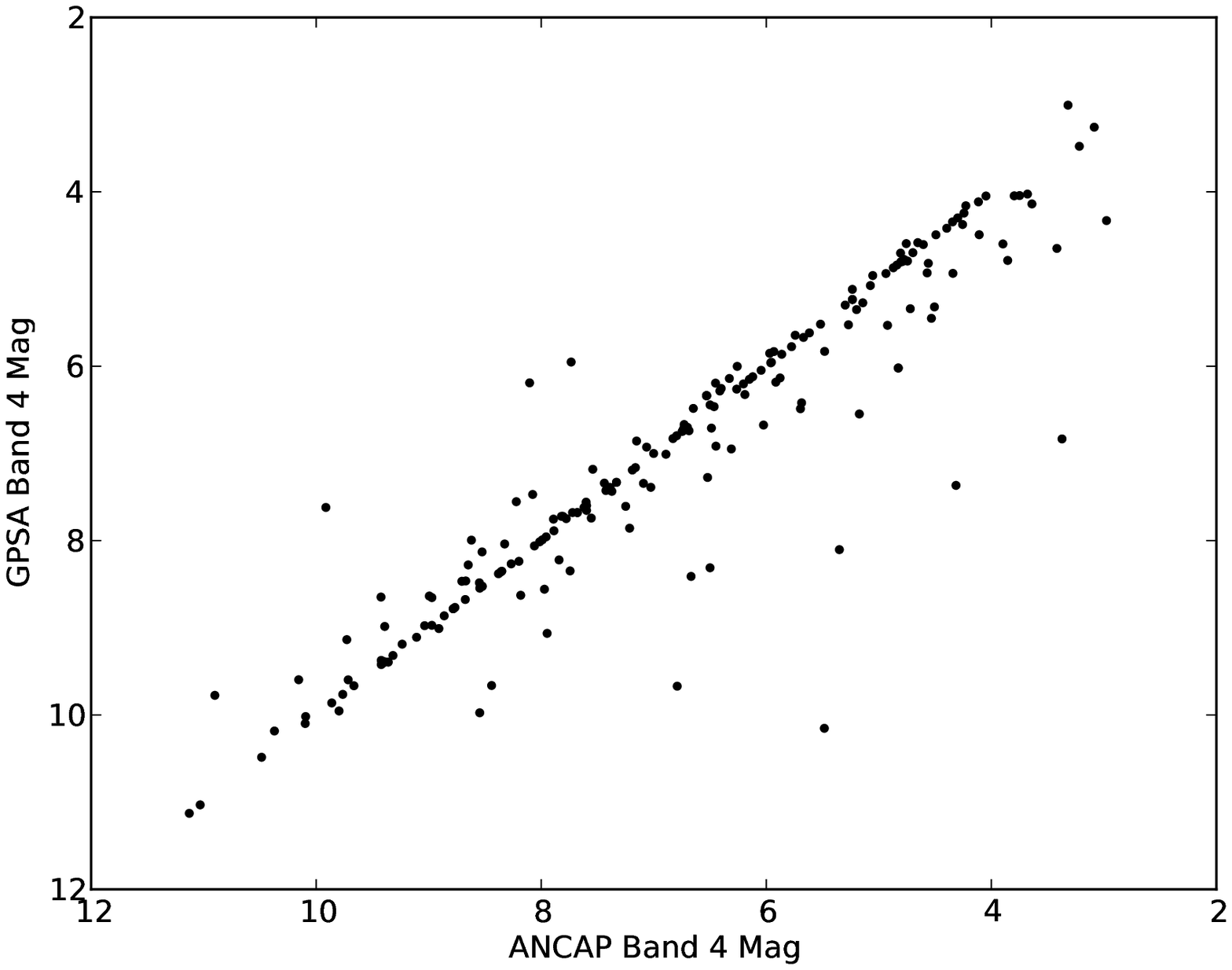} \\
\caption{The measured magnitudes of the common set of sources that have measurements from the GLIMPSE Point Source Catalogue and Archive (GPSC \& GPSA) and the Adaptive Non-circular Aperture Photometry method described in this paper (ANCAP). From top to bottom each panel refers to  3.6, 4.5, 5.8 and 8.0 $\mu$m respectively. In general the ANCAP measurements are consistent with, or brighter than, the GPSC and GPSA measured magnitudes.  This is as expected for the extended objects whose flux we recover via our aperture photometry technique whereas the point source measurements of the GPSC and GPSA underestimate the flux from extended objects. }
\label{Figure3}
\end{center}
\end{figure*}

In order to confirm the reliability of our measured aperture fluxes we compare against those sources that are also  contained in the GLIMPSE Point Source Catalogue (GPSC) and the GLIMPSE Point Source Archive (GPSA).  Within our sample of aperture photometry measurements there are $\sim$ 200 sources in common with the GPSC and $\sim$ 300 sources in common with the GPSA (the exact number in common varies from band to band). Plots of the aperture measured magnitudes (ANCAP) versus the GPSC and GPSA magnitudes at each waveband are shown in Figure~\ref{Figure3}. For the majority of sources in common with the GPSC and GPSA we measure magnitudes that are in either close agreement or up to a few magnitudes brighter than tabulated in the Catalogue or Archive (note that the typical magnitude errors in the GPSC and GPSA are $\le$ 0.2 mag). The latter sources are consistent with those that our technique  identifies as extended objects, hence the GPSC and GPSA have underestimated the total flux of these objects. We also identify a few objects ($\le$10 in total) within the common sample where our aperture photometry measurements are \emph{fainter} than those from the GPSC or GPSA. In general these objects lie within regions of complex background and, as our background subtraction method differs substantially from that used by  GLIMPSE (which determines the background via annular regions using DAOPHOT\footnote{See \url{http://www.astro.wisc.edu/glimpse/glimpse_photometry_v1.0.pdf} for more details on this process.}), it is likely that the discrepancy in measurements is caused by different background subtraction. In any case the fraction of sources affected in this way is small ($\le$5\%) and so we are confident that our aperture photometry technique yields good results for the majority of sources.

\

\subsection{Cross-matches with other catalogues: RMS, EGO and infrared clusters}
\label{sect:cats}

Earlier lower angular resolution investigations overlap extensively with the GLIMPSE/MMB region studied here. In particular, there has been intensive study of red objects selected from the MSX survey \citep{Lum02}, which has yielded the Red MSX Source (RMS) survey catalogue \citep[e.g.][]{urquhart2008,urquhart2011}. The RMS catalogue is the result of a Galaxy-wide search for Massive Young Stellar Objects (MYSO) and contains over 2000 objects that have been classified as either  Young Stellar Objects (YSO), compact or ultracompact HII regions, OH/IR stars or evolved stars based on their IR morphology, free-free radio emission, distance and luminosity. The RMS catalogue and its construction are detailed in \citet{urquhart2008,urquhart2011} and available at \texttt{http://ast.leeds.ac.uk/RMS}. 

A cross-match of MMB masers with the RMS catalogue has been made to 
identify the subset of MMB sources categorized in the RMS catalogue as YSOs or HII regions. The properties of even a small sample of these well-studied objects can 
then yield insights as to the likely nature of the less-well-studied majority of MMB sources. Within the region covered by both the MMB and RMS surveys there are 983 RMS objects of HII region or YSO classification and 673 MMB masers. Note that the RMS does not cover the inner region of the Galaxy ($-10\degr<l<10\degr$) and the MMB catalogue does not currently extend beyond longitude $20\degr$. Although the RMS objects were discovered by MSX colour selection and initially had a positional accuracy of $\sim$18\arcsec, all of the catalogue objects have since had more accurate positions determined from GLIMPSE or by dedicated near- or mid-IR imaging \citep{mottram2007,urquhart2008}. We thus selected a cross-matching radius of 2\arcsec\ to identify RMS objects associated with MMB masers.

EGOs are extended bright objects with excess emission at 4.5 $\mu$m (IRAC band 2), often coloured green in three colour IRAC images \citep{Cyg08}. They are thought to identify emission from shocked molecular gas via the presence of either CO bandhead or shocked H$_{2}$ lines \citep{Cyg09}, with shocked H$_{2}$ emission recently confirmed in at least one object \citep{debuizer2010}. \citet{Cyg08} have compiled a visually identified catalogue of EGOs drawn from the GLIMPSE survey, with later VLA follow-up showing a high incidence of these EGOs with Class II 6.7 GHz methanol masers (18 out of 28 searched). Here, we perform the inverse of the \citet{Cyg09} search by identifying MMB masers associated with EGOs to test the assertion of \cite{Cyg09} on a much larger sample of masers. We identify EGOs associated with MMB masers by cross-matching against the \citet{Cyg08} catalogue using a matching radius of 2\arcsec.

The visual search process that we employed to find infrared clusters and cluster candidates might be expected to include objects that are merely asterisms rather than bona fide clusters and to assist in the identification of clusters we also included a positional cross-match with catalogues of known clusters.
The cluster catalogues of \citet{Mercer}, \citet{Bica} and \citet{Froebrich} were used, plus the the UKIDSS DR4 cluster catalogue of 477 clusters  (Lucas et al.,  in preparation). To take into account the typical angular diameter of the infrared clusters our matching radius was set to 1\arcmin.

\section{Results}
\subsection{Visual Inspection and catalogue cross-matching}
\label{sect:class}

\begin{figure}
\includegraphics[scale=0.4]{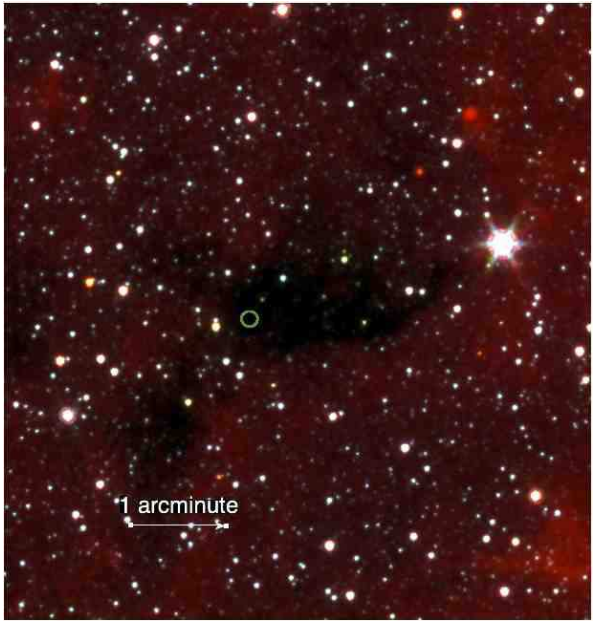} 
\caption{IRAC RGB image of a maser counterpart, with red, green and blue being bands 4, 2 and 1 respectively. The maser position marked with the green circle is located within an IRDC and  has no observable IRAC counterpart. }
\label{Figure4}
\end{figure}

\begin{figure}
\includegraphics[scale=0.4]{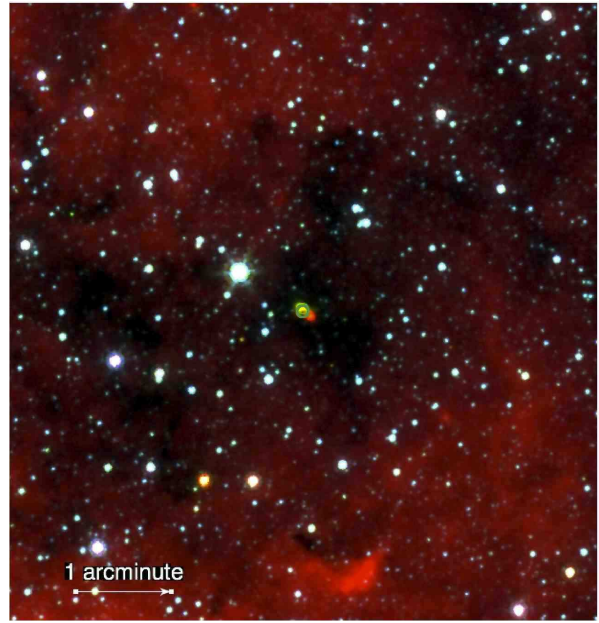} 
\caption{IRAC RGB image of a maser counterpart, with red, green and blue being bands 4, 2 and 1 respectively. The maser position is marked with the green circle. It is located within an IRDC but unlike Figure \ref{Figure4} it has a clear infrared counterpart.  }
\label{Figure5}
\end{figure}

\begin{figure}
\includegraphics[scale=0.53]{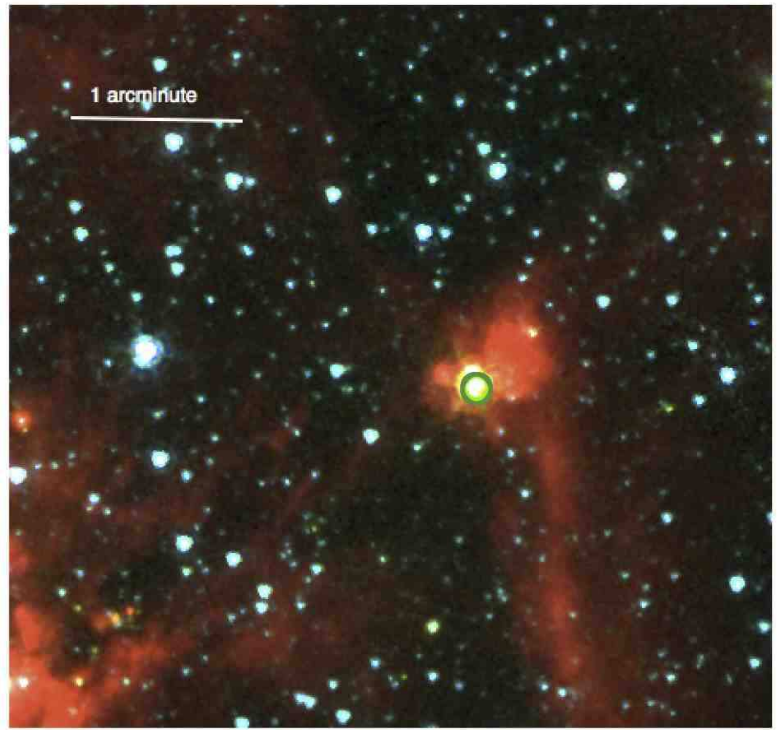} 
\caption{IRAC RGB image of a maser counterpart, with red, green and blue being bands 4, 2 and 1 respectively. The maser position is marked with the green circle. It can clearly be seen that the IR counterpart is extended in this case.}
\label{Figure6}
\end{figure}

Of the 776 masers within the GLIMPSE I, II \& 3D survey region, seven masers  could
not be used for the visual inspection process due to the maser falling slightly outside the  coverage of individual images.  For the  remaining 769  masers we find that they are located in one of the four following broad categories: 

\begin{enumerate}
\item Masers embedded within an IRDC with no IRAC counterpart. Hence
  they are infrared-dark at IRAC wavelengths due to extinction (or the
  lack of infrared emission). This represents 5\% (37) of our maser
  counterparts. Figure \ref{Figure4} shows a typical example of this category.

\item Masers that are located within, or are located on the perimeter of an IRDC but have a visible IRAC
  counterpart and therefore are infrared-bright.  This class
  represents 21\% (164) of our sample of maser counterparts. Figure \ref{Figure5} shows a
  typical example of this category.

\item Masers that are infrared-bright and  \textit{not} located within an
 IRDC. This class contains 62\% (473) of our sample. Figure
  \ref{Figure6} shows a typical example.

\item Masers that are infrared-dark (they have no identifiable counterpart in any band) but are \textit{not} located within an Infrared Dark Cloud. Our examination indicates that 12\%(95)
  of our objects are of this type.  
\end{enumerate}

We identify a total of 112 EGOs associated with 608  MMB masers located within the GLIMPSE I survey area (note that the \citealt{Cyg08} catalogue is comprised of EGOs found only in the GLIMPSE I images) that had good imaging data. All associated EGOs were  found as a result of positional matching against the \cite{Cyg08} catalogue. The visual search described in Sect.~\ref{sect:method} did not identify any more EGOs than the corresponding visual search of \cite{Cyg08}, suggesting that the \cite{Cyg08} catalogue of 304 EGOs is largely complete (at least towards 6.7 GHz masers). Our results imply that $\sim$18\% of masers are associated with EGOs, with the implication that  EGO-targeted searches would only be able to find $\sim$18\% of the masers. Conversely, we find that the fraction of EGOs that have a 6.7-GHz maser 
counterpart is 37\%. but varies according to the \cite{Cyg09} 
EGO classification as 'likely to be a massive YSO outflow' or 'possibly'. 
With this distinction, we find 25\% (41/165) of the 'possible' EGOs are associated with an 
MMB maser, increasing to 52\% (69/133) for 'likely' EGOs.  The latter 
statistic is in agreement with a small sample investigation by \cite{Cyg09}, yielding 18 methanol masers towards 28 'likely' 
outflow EGOs, although their result was biased by the EGO sample 
preferentially containing several known masers.


The combined infrared cluster catalogue used to cross match with the MMB
contains 118 clusters  and cluster candidates within the region covered by
our maser sample. It is found that eight masers are located within
one arcminute of these clusters.  Our cross matching with the early results of the 
UKIDSS Cluster search (Lucas et al., in prep.) shows that only six more MMB
masers, out of 306 that are currently covered in the region examined,
are within 1 arcminute of a cluster or cluster candidate. Thus the total number of masers that are found within  1\arcmin\ of an infrared clusters (or candidates) is 14.

We find a total of 82 MMB masers within 2\arcsec\ of an RMS object. Within the overlapping region common to both MMB and RMS surveys there are 983 RMS objects (of type HII region or YSO) and 673 masers, thus the detection fractions of maser-associated RMS objects and RMS-associated MMB masers are 8\% and 12\% respectively. Of the 82 MMB masers matched to an RMS object 56 are associated with type ``YSO or YSO?'', 22 with ``HII regions'', 2 with ``UC HII regions'' and 2 with ``HII/YSO'' blends. We did not find any associations between MMB masers and the evolved star categories in the RMS survey.



\subsection{Adaptive Non-circular Aperture Photometry}
\label{sect:ancap}

As previously mentioned, there are 769 masers from the MMB survey with available GLIMPSE images in all 4 wavelength bands.  Using the method described in Sect.~\ref{sect:ancap_method} we successfully managed to obtain aperture photometry in all four GLIMPSE bands toward a total of 512 masers. The remaining 257 masers either had no detectable counterpart to a level of 5$\sigma$ (as in categories \emph{i)} and \emph{iv)} in Sect.~\ref{sect:ancap_method}, a total of 132 masers) or were in regions too confused to reliably identify an infrared counterpart in all four bands (a total of 125 masers).  A simple positional association with the GLIMPSE Point Source Catalogue (GPSC) or Archive (GPSA) results in a total of 219 and 253 masers respectively with infrared counterparts measured at all four wavelength bands (see  Sect.~\ref{sect:method}). Hence our adaptive non-circular photometric method more than doubles the number of masers with known infrared counterparts measured at all four bands over catalogue-based searches.

We list the measured magnitudes for  a representative sample of the infrared counterparts to the MMB masers in Table \ref{tbl:ancapmags}. The full version of this Table is contained in the Online Supplement and comprises infrared counterparts to 626 masers published in \cite{Caswell2009}, \cite{Caswell2010}, \cite{green2010}, \cite{caswell2011} and \cite{green2012}. We will publish the magnitudes of the remaining infrared counterparts simultaneously with the positions and fluxes of the 6.7 GHz masers (Fuller et al.~2013, in prep). All statistics and figures of the infrared counterparts presented in this paper refer to the full interim catalogue of 769 masers.

\begin{table*}
\begin{minipage}{126mm}
\caption{Adaptive Non-Circular Aperture  Photometry (ANCAP) magnitudes for a representative sample of 20 MMB 6.7 GHz masers. The full version of this table is contained only in the online version of the paper. An ellipsis indicates that we were unable to measure an infrared counterpart to the maser at the appropriate wavelength, as discussed in Section \ref{sect:ancap_method}. The Morphology Class column refers to the four classes discussed in Section \ref{sect:class}. Where it was not possible to classify the morphology of the emission the entry is blank.}
\label{tbl:ancapmags}
\begin{tabular}{|l|l|l|l|l|l|l|l|}
\hline
  \multicolumn{1}{|l|}{Source name} &
  \multicolumn{2}{c|}{Equatorial coordinates} &
  \multicolumn{4}{c|}{ANCAP Magnitudes} &
  \multicolumn{1}{l|}{Morphology} \\
  \multicolumn{1}{|c|}{($l$, $b$)} &
  \multicolumn{1}{c|}{RA (2000)} &
  \multicolumn{1}{c|}{Dec.~(2000)} &
  \multicolumn{1}{c|}{3.6 $\mu$m} &
  \multicolumn{1}{c|}{4.5 $\mu$m} &
  \multicolumn{1}{c|}{5.8 $\mu$m} &
  \multicolumn{1}{c|}{8.0 $\mu$m} &
   \multicolumn{1}{l|}{Class}\\
  \multicolumn{1}{|c|}{(\degr \degr)} &
  \multicolumn{1}{c|}{(h m ss.s)} &
  \multicolumn{1}{c|}{(\degr \arcmin \arcsec)} &
  \multicolumn{1}{c|}{} &
  \multicolumn{1}{c|}{} &
  \multicolumn{1}{c|}{} &
  \multicolumn{1}{c|}{} &
  \multicolumn{1}{c|}{} \\
\hline
 312.108+0.262 & 14 08 49.31 & -61 13 25.1 &11.2&9.5&9.3&9.7& 3\\
  312.307+0.661 & 14 09 24.95 & -60 47 00.5 &9.9&8.6&7.4&5.9& 3\\
  312.501-0.084 & 14 12 48.95 & -61 26 03.2 &10.7&9.8&8.0&6.7& 3\\
  312.597+0.045 & 14 13 14.35 & -61 16 57.7 &11.5&9.8&9.0&8.4& 3\\
  312.598+0.045 & 14 13 15.03 & -61 16 53.6 &11.6&10.0&8.5&7.3& 3\\
  312.698+0.126 & 14 13 49.85 & -61 10 24.1 &12.5&10.6&9.5&8.5& 3\\
  312.702-0.087 & 14 14 25.12 & -61 22 29.0 &12.7&10.6&9.3&8.8& 3\\
  313.469+0.190 & 14 19 40.94 & -60 51 47.3 &10.6&7.6&6.1&5.3& 3\\
  313.577+0.325 & 14 20 08.58 & -60 42 00.8 &9.6&8.1&6.5&5.3& 3\\
  313.705-0.190 & 14 22 34.74 & -61 08 26.8 &10.5&8.1&6.4&5.7& 3\\
  313.767-0.863 & 14 25 01.73 & -61 44 58.1 &8.1&6.4&5.5&4.8& 3\\
  313.774-0.863 & 14 25 04.80 & -61 44 50.3 &\ldots &\ldots &\ldots &\ldots & 2\\
  313.994-0.084 & 14 24 30.78 & -60 56 28.3 &13.5&10.9&10.2&9.4& 3\\
  314.221+0.273 & 14 25 12.89 & -60 31 38.4 &8.9&7.1&5.7&3.7& 3\\
  314.320+0.112 & 14 26 26.20 & -60 38 31.3 &5.8&4.5&2.0&1.9& 3\\
  315.803-0.575 & 14 39 46.46 & -60 42 39.6 &10.2&8.8&7.1&6.3& 3\\
  316.359-0.362 & 14 43 11.20 & -60 17 13.3 &\ldots &\ldots &\ldots &\ldots & 3\\
  316.381-0.379 & 14 43 24.21 & -60 17 37.4 &11.0&10.0&8.1&6.6& 3\\
  316.412-0.308 & 14 43 23.34 & -60 13 00.9 &12.8&11.3&9.5&8.0& 3\\
  316.484-0.310 & 14 43 55.37 & -60 11 18.8 &12.7&9.7&7.6&6.4& 3\\
  316.640-0.087 & 14 44 18.45 & -59 55 11.5 &\ldots &\ldots &\ldots &\ldots & 3\\
   \hline\end{tabular}
\end{minipage}
\end{table*}

\begin{figure*}
\includegraphics[scale=0.3]{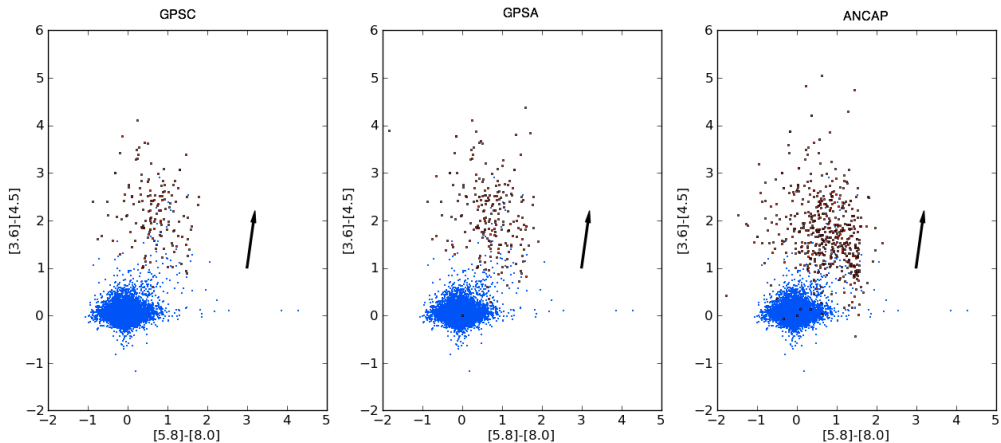} 
\caption{ [5.8]-[8.0] v [3.6]-[4.5] colour plot of GPSC, GPSA and ANCAP infrared counterparts respectively. This illustrates the broadening of colour space with increased source numbers and the general consistency of colour space between the three datasets. Reddening vector $A_{k}=10$}
\label{fig:colcol}
\end{figure*}

In Figure \ref{fig:colcol}  we show [3.6]$-$[4.5] versus [5.8]$-$[8.0] colour-colour plots of the maser counterparts compared to a sample population of 15\,000 sources  drawn randomly from the GLIMPSE Point Source Catalogue (from those sources with detections at all 4 bands). We show the colours of maser counterparts drawn from the GLIMPSE Point Source Catalogue (GPSC), GLIMPSE Point Source Archive (GPSA) and the ANCAP photometry measured here. A reddening vector of $A_{K}=10$ is displayed, and was calculated based on that of \citet{Gutermuth} and references therein. The maser counterparts are, as expected for YSOs, much redder than the general stellar population and show colours consistent with the smaller sample of 6.7 GHz masers investigated by \citet{Ell06}. All three photometric systems show good agreement in the colours of the maser counterparts, although the ANCAP counterparts occupy a marginally wider range in colour space than the GPSC or GPSA sample.

\begin{figure*}
\includegraphics[scale=0.6]{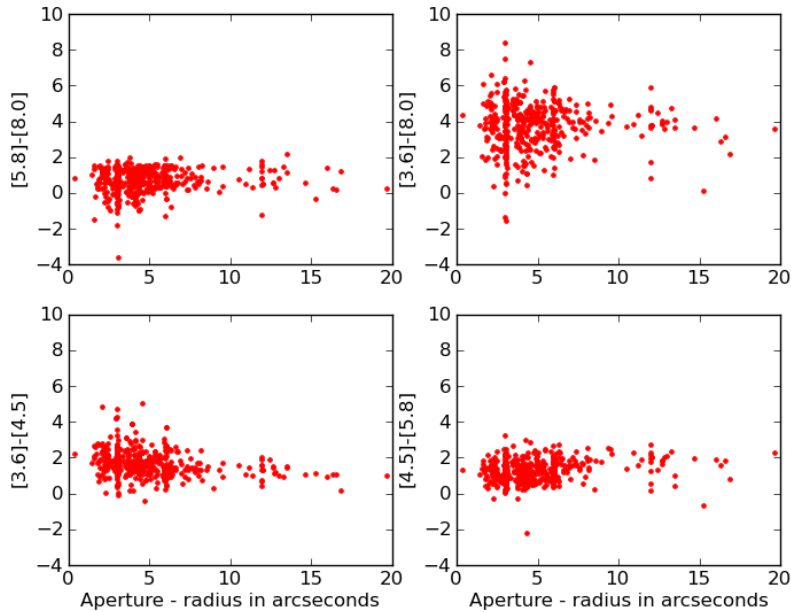} 
\caption{ Colours of the maser infrared counterparts as a function of aperture size used in the ANCAP process. }
\label{fig:apercol}
\end{figure*}

We investigate trends in colour with aperture size in Figure~\ref{fig:apercol}, where the aperture size is defined as the effective radius of the non-circular aperture (i.e. the radius of a circle with the same area as the non-circular aperture). In general the infrared colours of the counterparts display no marked trend with aperture size, although there is a weak  trend toward a bluer colour with increasing aperture in the [3.6]$-$[4.5] colour. This may be due to increased contamination by fore- or background stellar objects as the aperture size increases. There appears to be considerably larger scatter in the colours involving the 8.0 $\mu$m band, which could result from a contribution from extended PAH emission to the counterpart fluxes. 

\begin{figure*}
\includegraphics[scale=0.3]{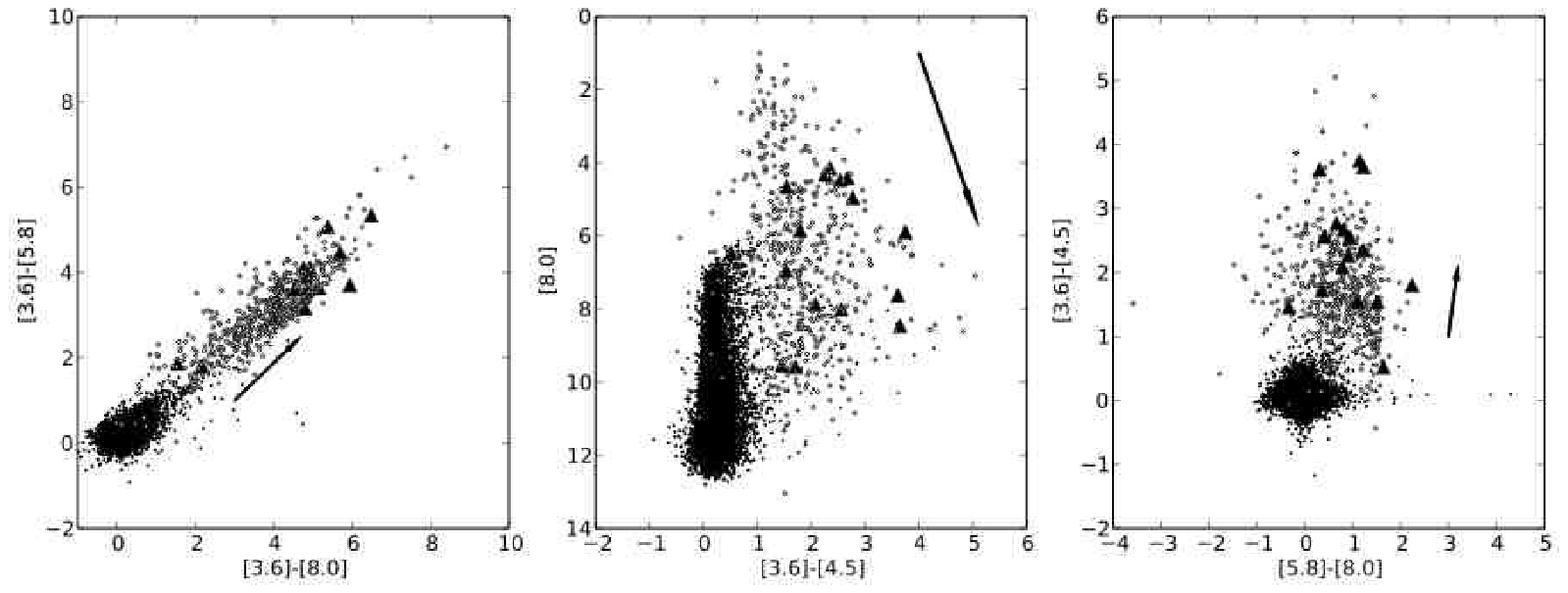} 
\caption{Colour-Colour and Colour-Magnitude plots showing the general GLIMPSE field population (dots), the maser counterparts  measured using our ANCAP method (open circles) and the maser counterparts from \citet{Ell06} (solid triangles). The reddening vector indicates $A_{k}=10$.}
\label{fig:colcol_ell}
\end{figure*}

The largest  sample of masers investigated in the GLIMPSE wavelength bands prior to the MMB Survey was that of \citet{Ell06}, who identified 41 Class II 6.7 GHz masers (drawn from  a statistically complete sample) to have counterparts in the GPSC and GPSA.  Figure~\ref{fig:colcol_ell} shows    [3.6]$-$[8.0] versus [3.6]$-$[5.8] and [5.8]$-$[8.0] versus [3.6]$-$[4.5]  colour-colour diagrams, plus a [3.6]$-$[4.5] versus [8.0] colour-magnitude diagram (c.f.~Figs.~15--19 of \citealt{Ell06}) of the  512 maser ANCAP infrared counterparts compared to the Ellingsen sample. It can be seen in this Figure that there is a close correspondence in colour-colour and colour-magnitude space between the masers whose infrared counterparts were determined by our ANCAP method and the GPSC/GPSA sample of \citet{Ell06}. The close correspondence between the two samples lends weight to the overall accuracy of the ANCAP procedure.
While the bulk of our ANCAP counterparts lie in the same colour-colour and colour-magnitude space as the Ellingsen sample, there are a small number of counterparts with bluer or redder colours. The counterparts with blue colours lie in the predominantly stellar region traced by the random GLIMPSE PSC sample (the blue dots in Figure~\ref{fig:colcol_ell}) and as such are likely to be foreground stellar contaminants.

\section{Discussion}

Here we draw together the results of our visual inspection of the GLIMPSE images, our adaptive non-circular photometry of extended IR counterparts to the MMB masers, and the results of the catalogue cross-matches. We dwell upon three main issues: the environments of the MMB masers, the colours of their infrared counterparts and the role of the masers as tracers of star formation.

\subsection{The infrared environments of MMB masers}

From our visual inspection of the GLIMPSE images we see that 6.7 GHz
methanol masers appear to occur in one of two environments: those that are embedded within
an infrared dark cloud (IRDC) and those that are associated with bright, often extended, mid-infrared emission. 
We also observe an apparent intermediate stage whereby there is a bright mid-IR counterpart associated with the maser, in turn embedded within an IRDC.
We find 5\% of our MMB maser sample to be associated with IRDCs, 62\% of the sample to be associated with bright mid-IR emission and a further 21\% to be associated with mid-IR emission embedded within IRDCs. The remaining 12\% of the MMB maser sample are not associated with any mid-IR emission or visible IRDCs within the GLIMPSE images.

The difference in the environments traced by the masers indicated that 6.7 GHz masers  trace more than one common evolutionary stage, as suggested by \cite{Ell06} who also found a similar fraction of 6.7 GHz masers (albeit from a much smaller sample) to be associated with IRDCs. If we assume that the different mid-IR environments reflect a common evolutionary theme, then a plausible scenario is that the IRDC-associated and infrared-dark masers trace the earliest and most deeply embedded phase of star formation, followed by a transitional stage whereby the embedded source begins to break free of the IRDC, becoming visible in the mid-IR, and finally followed by a bright extended mid-IR phase where the source exciting the maser has disrupted the IRDC. 
Many of the bright mid-IR sources are found to be  extended with respect to the GLIMPSE point spread function and at least some of these sources correspond to compact HII regions rather than the point-like morphologies expected by ultracompact or hypercompact HII regions.

However, this hypothesis does not take into account the distance of the maser -- for example beyond a distance of 10 kpc a typical ultracompact HII region would be unresolved in GLIMPSE. Accurate distance determinations (free from kinematic distance ambiguities) are required to determine the spatial extent of the extended mid-IR emission combined with follow-up radio observations to confirm the HII region hypothesis by detecting free-free emission that is morphologically associated with the mid-IR emission \citep[e.g.][]{hoare2007,mottram2007,urquhart2008}

To confirm that the IRDC-associated and infrared-dark masers trace an embedded phase of star formation within an IRDC requires longer-wavelength observations, e.g.~from the MIPSGAL survey \citep{carey} or the Hi-GAL survey \citep{molinari2010b}. A study based on the results of the MIPSGAL and Hi-GAL surveys is in preparation and preliminary results indicate that these masers do indeed trace embedded star formation within these clouds. In addition, if these masers represent early deeply embedded phases of massive star formation some of them may show detectable hypercompact HII regions, \citep[e.g.][]{Longmore07}. High frequency high resolution radio continuum follow-up observations will be made to discover which masers are in this category. The role of the infrared-dark masers that are not associated with IRDCs is much less clear. We hypothesise that these masers represent distant sources whose associated IRDCs have either been rendered invisible by foreground diffuse Galactic mid-IR emission, or are too small or low contrast to be effectively detected by GLIMPSE.

\subsection{The colours of the mid-infrared counterparts of MMB masers}

Prior to the work carried out in this paper, the largest investigations
of the GLIMPSE counterparts of 6.7\;GHz masers have been carried out
by \cite{Ell06,Ell07,breen2010b} and \cite{Breen2011}. In summary, \cite{Ell06}
took a sample of fifty six 6.7\;GHz methanol masers and found that 29 
masers (52\%) have GPSC counterparts within 2\arcsec. This leads to a
colour-magnitude selection criterion of [3.6]-[4.5]$>$1.2 and
[8.0]$<10$. In a follow up paper \citep{Ell07} this criterion was used
to select 5676 GPSC objects. Of these the 100 brightest in 8\;$\mu$m
and the 100 reddest at [3.6]$-$[4.5] were selected as candidates for
radio observations in order to detect 6.7 GHz methanol
masers. This led to the detection of 38 maser sources, of which nine
were new discoveries.

\cite{Breen2011} followed up 580 6.7 GHz masers drawn from the MMB catalogue at 12.2 GHz with the Parkes telescope and identified 12.2 GHz maser counterparts towards a total of 250 masers. \cite{Breen2011} also investigated the detection statistics and colours of their maser sample with the GLIMPSE point source catalogue in the same manner as \cite{Ell06}. The detection statistics of the \cite{Breen2011} masers  are in agreement with those of \cite{Ell06}, although there is marginally higher detection rate for 6.7 GHz masers without associated 12.2 GHz emission than for sites showing both maser transitions. \cite{Breen2011} interpret this difference as an indication that sites with both maser transitions may be more evolved than those showing only 6.7 GHz maser emission. The infrared colours of the maser counterparts were found to be similar to the smaller sample of \cite{Ell06}.

In our visual inspection of the GLIMPSE images and positional cross-matches with the GPSC and GPSA we found that a significant fraction of the maser counterparts (480 masers out of 769) are extended in the mid-IR. As described in Sect.~\ref{sect:ancap} we measured the fluxes in all 4 IRAC bands of 512 MMB maser counterparts using our adaptive non-circular aperture phometry method, which is $\sim$ a factor 2 increase over simply matching against the GPSC or GPSA catalogues, an order of magnitude increase over the sample of \cite{Ell06} and more than doubling  the \cite{Breen2011} sample . Despite the much larger size of our sample we find a striking agreement between the colours of our maser counterparts and those of \cite{Ell06} and \cite{Breen2011}. Figure~\ref{fig:colcol_ell} reveals that both samples occupy a very similar colour space and are much redder than the typical GLIMPSE population. In our larger sample we observe a marginally wider scatter in colours than \cite{Ell06} and \cite{Breen2011}, particularly in colours involving the 8 $\mu$m band, however there are no clear trends. A small number of masers show stellar colours consistent with the general GLIMPSE population. These masers are likely to be chance line-of-sight alignments with foreground or background stars.

As noted
in \cite{Ell06} the masers appear to occupy a colour-space similar to
that of Class 0 protostars as modelled in \cite{Whitney03}. However,
it is also similar to the region occupied by HII regions as shown in
\cite{Cohen07} and we should not draw the conclusion that the
maser counterparts invariably represent a Class 0-like object with an in-falling
envelope without first excluding the possibility that some, at least, are
HII regions.

\subsection{The relationship between maser and infrared properties}

With our large sample of maser infrared counterparts and the measured properties (e.g.~6.7 GHz flux, luminosity, velocity range) of their corresponding masers from the MMB Survey we are in the position to search for correlations between properties of the infrared and 6.7 GHz maser emission. This would enable us to search for potential evolutionary effects similar to those suggested by \citet{breen2011b} and \cite{breen2012} in which the  luminosity and velocity range of water and methanol masers are found to be correlated with the volume-averaged density of their host molecular clumps. We searched for trends in all of the measured IR properties contained in this paper (magnitude, colour, morphology) against each of the measured 6.7 GHz properties from the MMB survey (6.7 GHz flux and velocity range). We found no significant trends in any case. There is no tendency for the MMB masers to have measurably  different properties for different morphological classes of infrared emission (e.g.~IRDCs or Extended Green Objects), nor for different mid-infrared colour or magnitude.  We cannot rule out evolutionary effects within our sample, although we can conclude that the physical mechanism behind the GLIMPSE emission is likely to be unrelated to that driving the masers themselves.

\subsection{Methanol masers as tracers of star formation}
\label{sect:crossmatch}

The MMB masers are found to be rather weakly associated with Extended Green Objects (EGO) and Red MSX Survey (RMS) sources, with $\sim$18\% of MMB masers found to be associated with EGO and 12\% found to be associated with RMS HII regions or YSOs (with 20 masers in common between EGO and RMS, which is 3\% of the total number of masers).  We find that the EGO detection rate is consistent with that of \cite{Cyg08} when the split between ``likely'' and ``probable'' EGOs is considered. Accounting for masers that are associated with both EGOs and RMS objects, there are 23\% of the MMB sample that are found to be associated with either an EGO or an RMS object. A further 17\% of MMB masers have no detectable mid-infrared counterparts, either being associated with an IRDC or with no detectable counterpart, which means that 60\% of the MMB sample are infrared-bright but are \emph{not} associated with other known massive star formation tracers such as EGO or RMS objects. 

The question thus arises as to why  the RMS and MMB surveys are only weakly correlated, if the hypotheses that both types of object trace current massive star formation are correct? To consider this question we must first recall that RMS is a survey that is primarily based on infrared selection, with specific MSX colour criteria required for selection as a candidate massive YSO \citep{Lum02} followed by detailed inspection at other wavelengths to weed out contaminants \citep{urquhart2008}. The MMB survey on the other hand does not rely upon infrared selection, merely upon the initial detection of a 6.7 GHz maser by Parkes and its successful recovery at higher angular resolution by the ATCA. So any difference between the populations traced by RMS and MMB may simply be due to a difference in the intrinsic mid-infrared colours of RMS sources and 6.7 GHz masers.

\begin{figure}
\includegraphics[scale=0.4]{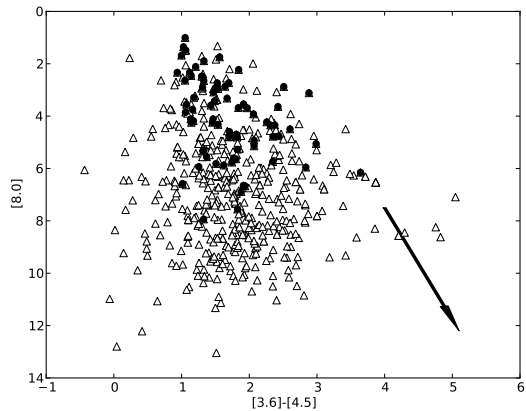} 
\caption{Colour - Magnitude diagram comparing masers with counterparts within RMS  (solid circles) to the GLIMPSE masers (open triangles). We show a reddening vector of $A_{k}=10$.}
\label{Figure10}
\end{figure}

We investigate the colours of the RMS and maser populations in Figure \ref{Figure10}, by plotting the $[3.6]-[4.5]$ colour versus the $[8.0]$ $\mu$m magnitude for maser counterparts both associated and unassociated with RMS sources. It can clearly be seen in Figure \ref{Figure10} that the RMS-associated masers are both brighter at  $[8.0]$ $\mu$m  and bluer in $[3.6]-[4.5]$  than those maser counterparts that are not associated with RMS objects. The former is relatively easy to explain due to the fact that all RMS sources were selected from the MSX Survey. The MSX Survey was much shallower than GLIMPSE and has a limiting magnitude of $\sim$6.2 at 8.2 $\mu$m \citep{cohen2000}, although may reach deeper in regions that are less confused by the diffuse Galactic background. Hence the non-association of MMB masers that are fainter than a magnitude of $\sim$6.2  at 8.0 $\mu$m with RMS sources is entirely consistent with the limiting magnitude of MSX. A total of 304 of the 512 MMB maser counterparts for which we obtained photometry via the ANCAP procedure are fainter than magnitude 6.2 at 8.0 $\mu$m.

\begin{figure}
\includegraphics[scale=0.35]{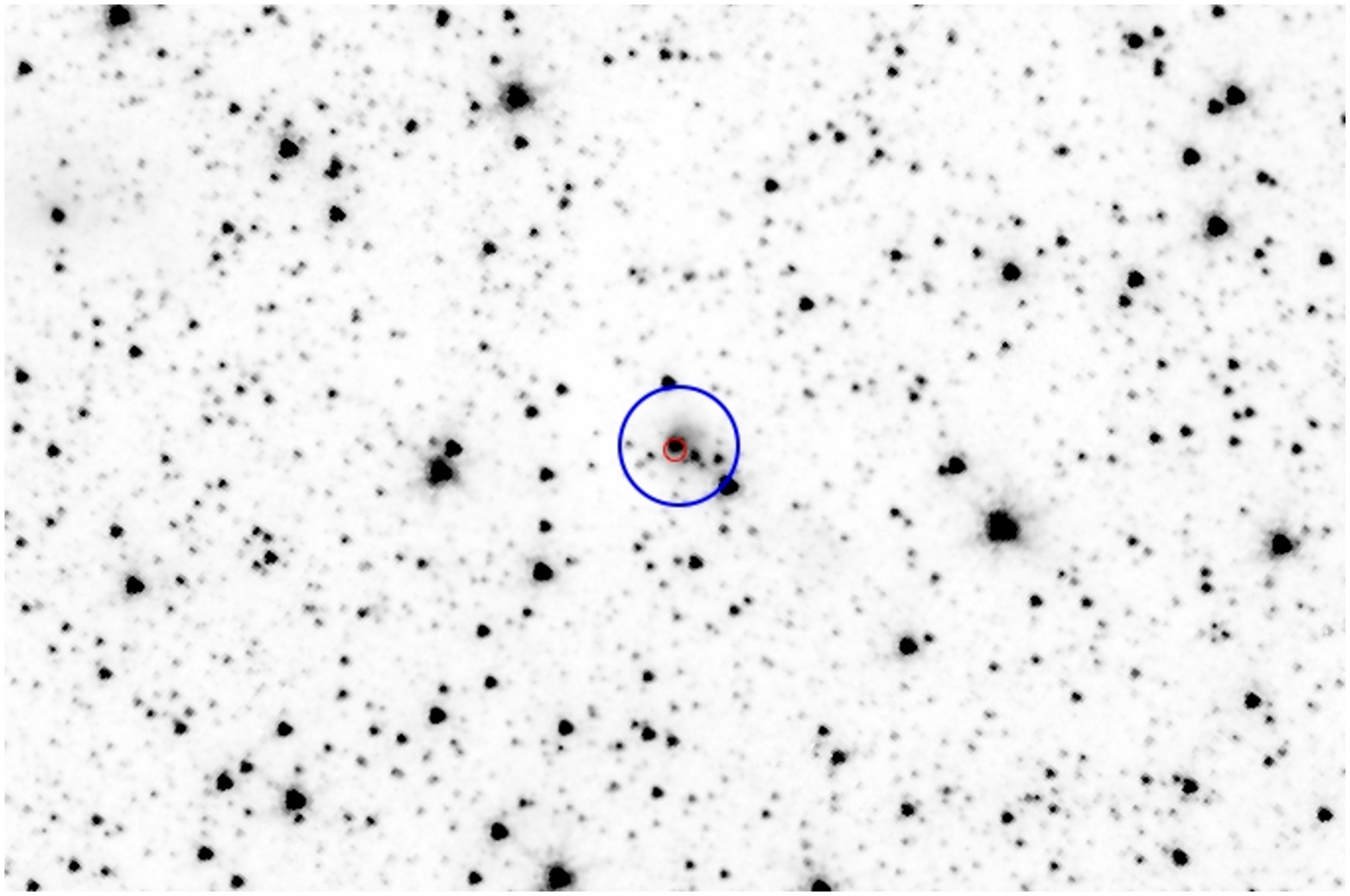}
\includegraphics[scale=0.35]{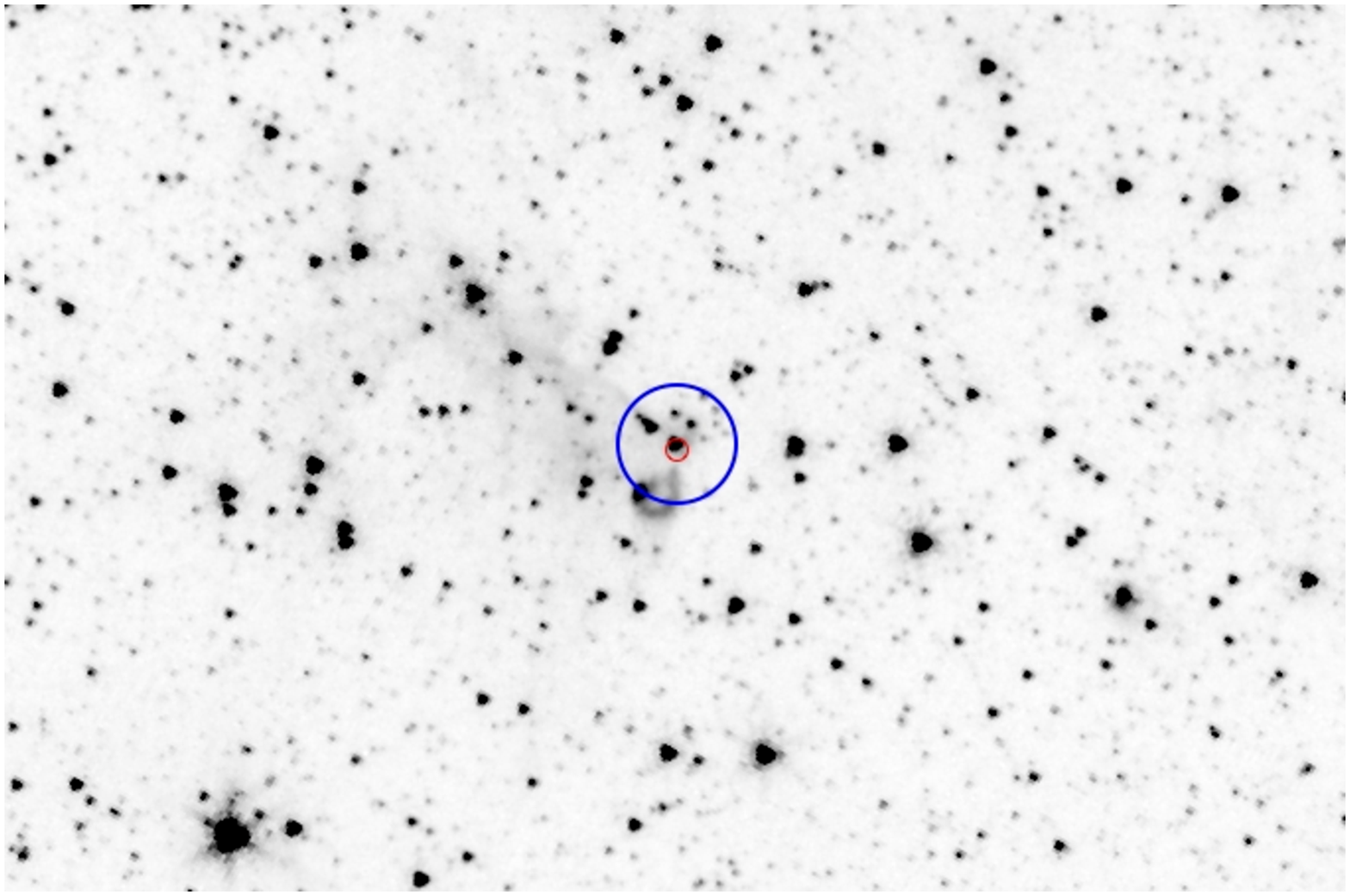}
\caption{GLIMPSE 8.0 $\mu$m images of two masers from the MMB catalogue that are not associated with RMS sources. The larger circle represents the beam size of MSX at 8 $\mu$m and the smaller circle indicated the position of the MMB maser. The GLIMPSE images clearly reveal a number of nearby confusing objects, possibly stellar in nature, whose flux results in the MMB maser infrared counterpart failing the RMS selection criteria.}
\label{fig:confusion}
\end{figure}

Secondly, it can be seen that the masers extend to a bluer region in $[3.6]-[4.5]$ than the RMS sources, extending into the domain where $[3.6]-[4.5] < 1$. 
The RMS sources are (by definition) selected to be red in MSX bands, with one key criterion being that $F_{21 \mu m} > F_{8 \mu m}$, corresponding approximately to $[8]-[21] > 0.8$. If this is similar to a corresponding criterion  $[3.6]-[4.5] > 1$ for the typical RMS source this could fully account for the observed absence of RMS sources with  $[3.6]-[4.5] < 1$ in Figure \ref{Figure10}. A secondary effect may arise from the poorer angular resolution of MSX relative to GLIMPSE for 8.0 $\mu$m measurements. We might then interpret the fact that masers trace bluer objects than the RMS as being due to contamination of the MSX photometry by nearby, potentially stellar, objects. In Figure \ref{fig:confusion} we show 8.0 $\mu$m GLIMPSE images of two MMB masers that are not associated with an RMS source. The large circle marked on the image represents the beam size of MSX at 8.0 $\mu$m and shows the presence of a number of additional objects within the MSX beam. The flux from these stellar objects increases the 8.0 $\mu$m MSX flux sufficiently that these two masers fail the $[8]-[14]$ selection criterion of the RMS Survey. We checked a random sub-sample of MMB masers using MSX photometry and found that this is generally the case for those masers with bright MSX counterparts. The RMS survey is thus naturally biased towards bright massive star forming regions that are isolated on the scale of the MSX beam.

Hence the reason that MMB and RMS appear to trace weakly related populations is entirely consistent with the properties of the initial RMS selection criteria, i.e. towards  brighter infrared counterparts with $[8.0] \stackrel{\sim}{>}$ 6.2 and for regions with relatively unconfused MSX photometry. The MMB survey thus offers a powerful technique for identifying massive star forming regions that is independent of infrared selection \citep[e.g.][]{ellingsen1996,ellingsen2005}.  

Looking in more detail at the classifications of RMS objects found in the MMB maser sample, we see that the majority of the RMS objects found to be associated with masers belong to the YSO category (56/82 objects), with the next highest category association being ``HII region'' (22/82 objects). This is consistent with our visual inspection of the GLIMPSE images where $\sim$50\% of the MMB maser sample were found to be point-like \citep[similar to the near-IR/\emph{Spitzer} morphology of YSOs found by][]{urquhart2008} and the remainder either extended or infrared-dark. 

The RMS survey were able to determine unambiguous kinematic distances, and hence luminosities, for 78 of the 82 RMS sources associated with MMB masers. All of these RMS sources have luminosities that are consistent with massive star-forming regions; the minimum luminosity found is 10$^{4}$ L$_{\odot}$, with a median for the sample of 3.2$\times$10$^{4}$ L$_{\odot}$. Although we are only able to determine the bolometric luminosity for a small fraction of MMB masers ($\sim$12\%),  by this approach we are confident that our sample of masers does not contain a significant fraction of low luminosity objects. The RMS survey contains YSOs with luminosity down to a few tens to hundreds $L_{\odot}$ and so, if low-luminosity YSOs were common in the MMB sample we would expect to have discovered some of them in our catalogue cross-match.

The implication is that 6.7 GHz masers  trace a range of evolutionary stages in massive star formation, from those so deeply embedded within their parent clouds that they are undetectable at 8 $\mu$m \citep[e.g.][]{Parsons09}, to YSOs and HII regions. Caution must, however, be exercised as the reality is that massive star forming regions are highly complicated and confused complexes. We note that increasing the matching radius for MMB masers and RMS objects from 2\arcsec\ to 5\arcsec\ almost doubles the number of matched MMB masers. A more sophisticated analysis, using maps of extended structures and precise point source positions, is needed to disentangle the precise relationship between 6.7 GHz masers, HII regions and YSOs. 

  In a future publication we will investigate the bolometric luminosity of the MMB sample by using \emph{Herschel} measurements of their SEDs from the Hi-GAL survey \citep{molinari2010b}. We are also conducting distance determinations  \citep[e.g.][]{green2011} and additional multiwavelength follow-ups (similar to the RMS survey) to complement Herschel measurements  so as to achieve our goal of extending full evolutionary classification to all MMB sites.


\subsection{The infrared cluster maser connection}
The cross match of the MMB and the 
infrared cluster catalogues (Mercer, Bica and Froebrich catalogue, plus the UKIDSS DR4 catalogue in preparation)  yielded only 14 cross matches. A visual inspection of the GLIMPSE images and of the UKIDSS
GPS \citep{Lucas08} images, where available, shows that the majority
of the maser-associated clusters are not distinct in the near infrared
and that only two masers are clearly associated with infrared
clusters. We also cross-matched the RMS catalogue to our cluster sample with the result that only 28 clusters
are associated with a RMS YSO or HII region. These two observations
tentatively suggest that massive star formation has ended by the time
a cluster becomes distinguishable in the near infrared. However, we
must consider the fact that the cluster catalogues used in this work contain
mostly nearby clusters \citep{Bica}, or clusters with predominantly low 
mass star formation, whilst the MMB and RMS sources are spread over a wide 
range of distances. Therefore we should be cautious in drawing any firm
conclusions from this match. Cross matching the MMB with the much
deeper UKIDSS Infrared cluster catalogue (Lucas et al in prep.),
should provide a more reliable indication due to the number of more
distant clusters detected; a preliminary cross match with the UKIDSS
Infrared cluster catalogue so far indicates that the lack of correlation 
between masers and infrared clusters continues within the deeper survey.

\section{Summary \& Conclusions}

We have carried out a detailed study of the mid-infrared environments of the 6.7 GHz methanol masers discovered in the MMB Survey using the \emph{Spitzer} GLIMPSE survey. Our study comprises 776 6.7 GHz masers within the GLIMPSE I, II and 3D survey regions. We have implemented an adaptive non-circular aperture photometry technique (ANCAP) that determines the mid-IR flux of infrared counterparts to the masers in all 4 GLIMPSE bands. Our ANCAP technique doubles the number of masers with fluxes in all four bands (512 masers) compared with the corresponding number of counterparts obtained from the GLIMPSE point source catalogue (219 or 253 masers depending on whether the reliable GLIMPSE Catalogue or complete GLIMPSE Archive are used). We also examine the positional association of the masers with a number of star formation tracers: EGOs, IRDCs, RMS sources and infrared clusters.

We reach the following conclusions:

\begin{enumerate}

\item Visual inspection of the images around each maser reveals a generally complex infrared morphology with the maser counterparts often being extended with respect to the \emph{Spitzer} PSF. The morphology of the maser environments falls into one of four broad categories: \emph{a)} the maser is embedded within an IRDC with no IRAC counterpart and thus has no detectable infrared counterpart (37 masers); \emph{b)} the maser is located within an IRDC and has a detectable counterpart (164 masers); \emph{c)} infrared-bright masers, often with extended counterparts, and that are not associated with IRDCs (473 masers); and \emph{d)} masers with no detectable infrared counterpart and that are not associated with an IRDC (95 masers).

\item We find that colours of the MMB maser counterparts sample agree very closely with those of a smaller sample studied by \citet{Ell06} and \cite{Breen2011}, but with a marginally larger scatter in colour. The region of mid-infrared colour space traced by the masers is similar to that of Class 0 protostars \citep{Whitney03} and HII regions \citep{Cohen07}.

\item We find 112 Extended Green Objects (EGOs) from the \citet{Cyg08} catalogue to be associated with MMB 6.7 GHz masers (out of 608 masers searched), with the implication
that EGO-targeted searches are able to detect only 18\% of the 6.7 GHz masers.
We also investigated extensively the fraction of EGOs that have a 6.7 GHz
maser counterpart, and show that the investigation by
\citet{Cyg09} is compatible with our more complete statistics once the \citet{Cyg09} sample is corrected for the ``likely'' rather than ``possible'' EGOs.

\item Few masers (and few embedded RMS sources) are found within 1\arcmin\ of infrared clusters which suggests that ongoing massive star formation has largely ceased by the time a cluster is discernible in the near-infrared. However this result may be due to a distance bias in the catalogue of infrared clusters used. 

\item The MMB masers are found to be rather weakly associated with RMS objects. Only 82 MMB masers lie within 2\arcsec\ of an RMS object of type ``HII region'' or ``YSO'', with the majority (56 masers) associated with type ``YSO''.  Combining this result with the EGO sample implies that 60\% of the MMB masers have a detectable infrared counterpart within the GLIMPSE survey but are \emph{not} associated with other known massive star formation tracers. We corroborate the common belief that 6.7-GHz masers are associated with 
an early stage of star formation, often extending earlier than the 
ultracompact HII region phase  \citep{Ell06,Walsh98}. Hence, the MMB survey offers a powerful way of identifying massive star forming regions that is independent of infrared selection \citep{ellingsen2005,ellingsen1996}.

\item MMB masers appear to trace a range of phases in the massive star formation process, with some masers associated with IRDC and some with infrared bright YSOs or HII regions.  Future work to investigate the nature of infrared-dark masers is ongoing and includes studies of the far-IR \& sub-mm SEDs \citep[using the \emph{Herschel} Hi-GAL survey,][]{molinari2010b} and of their high frequency radio continuum to identify hypercompact HII regions  \citep[e.g.][]{Longmore07}

\end{enumerate}

\section{Acknowledgments}
We would like to thank the referee, Andrew Walsh, for a thorough and constructive report that improved our paper. We wish to thank the University of Hertfordshire and the Science and
Technology Facilities Council for their support. This research has
made use of NASA's Astrophysics Data System Abstract Service; the NASA
/  IPAC Infrared Science Archive (which is operated by the Jet
Propulsion Laboratory, California Institute of Technology, under
contract with the National Aeronautics and Space Administration);
SAOImage DS9, developed by Smithsonian Astrophysical Observatory;
FunTools developed by  High Energy Astrophysics Division of the
Smithsonian Astrophysical Observatory and data products from the
GLIMPSE survey, which is a legacy science program of the Spitzer Space
Telescope, funded by the National Aeronautics and Space
Administration.

\bibliographystyle{mn2e} 
\bibliography{paper1w}

\label{lastpage}

\end{document}